\documentclass[%
preprint,
aip,
amsmath,amssymb,
aps,
]{revtex4-2}
\usepackage{dcolumn}
\usepackage{bm}
\usepackage{graphicx}
\usepackage{color}

\newcommand{\jgr}{J. Geophys. Res.}
\newcommand{\apjl}{Astrophys. J. Lett.}
\newcommand{\grl}{Geophys. Res. Lett.}

\newcommand{\ssr}{Space Sci. Rev.}
\newcommand{\aap}{A \& A}
\newcommand{\planss}{Plan. Space Sci.}

\newcommand{\crd}{\color{black}}
\newcommand{\cbk}{\color{black}}

\begin{document}
\title{Plasma Heating by Magnetic Reconnection and Alfv\'en Waves in Collisionless Shocks} 
\author{Masahiro Hoshino}
\affiliation{Department of Earth and Planetary Science, The University of Tokyo, Tokyo 113-0033, Japan.}
\email{mhoshino@g.ecc.u-tokyo.ac.jp}

\author{Shuichi Matsukiyo}
\affiliation{Department of Earth System Science and Technology, Kyushu University, Fukuoka 816-8580, Japan.}

\author{Takanobu Amano}
\affiliation{Department of Earth and Planetary Science, The University of Tokyo, Tokyo 113-0033, Japan.}

                       
\begin{abstract}
In collisionless shock waves, the ion and electron temperatures generally differ, and the mechanisms governing their energy partitioning remain one of the central unresolved problems in shock physics. Although energy can be exchanged through various plasma waves generated upstream and downstream of the shock, large‑amplitude Alfvén waves excited by ions reflected at the shock front play a particularly important role in high–Mach‑number shocks. 
In this study, we focus on these large‑amplitude Alfvén waves and examine not only plasma heating through conventional wave–particle interactions but also preferential ion heating driven by magnetic reconnection within current sheets formed by trains of such waves. Using particle‑in‑cell (PIC) simulations, \crd we find a correlation between ion heating and X-type magnetic field structures in the downstream region, and we suggest \cbk that magnetic reconnection and magnetic annihilation within these current sheets influence the energy partitioning between ions and electrons.
\end{abstract}

\maketitle
\section{Introduction}

Collisionless shock waves, where the mean free path among particles is far longer than the shock structure, play a central role in plasma heating and particle acceleration throughout most astrophysical environments. When a low entropy, supersonic upstream flow encounters the shock front, non-Maxwellian velocity distribution functions can rapidly develop. These non-Maxwellian distributions excite a broad spectrum of waves, ranging from low-frequency magnetohydrodynamic (MHD) modes to high-frequency electrostatic and electromagnetic fluctuations. Wave–particle interactions associated with these fluctuations drive the plasma toward relaxation into a thermal Maxwellian distribution. However, the ion and electron temperatures generally remain unequal, and the partitioning of energy between the two species continues to be one of the central unresolved problems in shock physics.

At the initial encounter of the upstream ion and electron flows with the shock front, where abrupt changes in the magnetic and electric fields occur, the supersonic plasma is rapidly decelerated, and individual particles become strongly scattered in phase space. In most cases, the ion “effective” temperature, which reflects the degree of velocity-space randomization, becomes substantially higher than the electron “effective” temperature. This disparity arises primarily because ions carry much larger bulk kinetic energy than electrons, as dictated by their mass ratio.

Farther downstream, the total plasma temperature of a collisionless shock can be estimated using the Rankine–Hugoniot (R-H) jump conditions. However, the individual ion and electron temperatures cannot be determined from the R–H relations of the conservation laws alone; understanding them requires detailed knowledge of shock dynamics and the underlying energy‑transfer processes. In high‑Mach‑number shocks, where the Mach number exceeds the so‑called critical Mach number \cite{Kennel85}, a fraction of the incoming upstream ions can be reflected at the shock front \cite{Paschmann81,Leroy81}. These reflected ions form a beam‑like component that serves as a source of various plasma waves in the shock upstream region, and these processes are well documented in in-situ spacecraft observations of planetary bow shocks \cite{Paschmann80,Bame80}. Such wave–particle interactions constitute one of the key channels for energy transfer between ions and electrons.

Extensive numerical simulations have also greatly advanced our understanding of collisionless shock dynamics \cite{Leroy81,Burgess12}, and especially kinetic particle-in-cell (PIC) simulations are indispensable for uncovering shock dynamics by surveying a wide range of shock parameters. Recent studies combining simulations and observations have placed particular emphasis on the partitioning of energy between ions and electrons, especially the ion‑to‑electron temperature ratio in supernova remnants with very high Mach number shocks and the solar wind with mildly high Mach number shocks \cite{Ghavamian13,Bohdan20a,Bohdan20b,Tran20,Raymond23}.  This research direction has benefited from the combined use of kinetic simulations and in‑situ or remote‑sensing observations, providing a more integrated perspective indicating that the ratio of electron temperature to ion temperature declines with increasing shock speed or Alfv\'en Mach number. Nevertheless, a comprehensive understanding of energy partitioning remains elusive, largely because it depends on a complex interplay of parameters—including Mach number, shock angle, the ion and electron plasma $\beta$, and waves and turbulence, among others.

In this study, we investigate the mechanisms of energy partitioning in collisionless parallel shock waves by employing both one‑dimensional and two‑dimensional particle‑in‑cell (PIC) simulations, where the upstream magnetic field is aligned with the plasma flow. We begin by examining the generation of large‑amplitude Alfvén waves excited by ions reflected at the shock front \cite{Giacalone97,Caprioli14}. Plasma heating is known to arise not only from these large‑amplitude waves \cite{Otsuka19} but also from broadband fluctuations produced through parametric decay and modulational instabilities \cite{Goldstein78,Hoshino89}. Beyond such wave–particle and wave–wave interactions, we place particular emphasis on plasma heating driven by magnetic reconnection within current sheets formed by trains of large‑amplitude Alfvén waves. Recent satellite observations and kinetic simulations have demonstrated that magnetic reconnection occurs in the Earth's bow shock \cite{Karimabadi14,Gingell19,Gingell23,Bessho23,Bessho25a,Bessho25b}. Moreover, in higher Mach number shocks relevant to supernova remnants, magnetic reconnection has been shown to develop within current sheets generated by the Weibel instability upstream of the shock \cite{Matsumoto15,Matsumoto17,Bohdan19a}. Heating associated with magnetic reconnection constitutes a key energy‑dissipation pathway that influences the ion–electron energy partitioning even in collisionless shock environments.

To quantify the role of magnetic reconnection in shock‑driven energy dissipation, we conduct both one‑dimensional and two‑dimensional PIC simulations in which reconnection can be selectively suppressed or enabled. By comparing these configurations, we assess the resulting ion–electron energy partitioning. Our results \crd suggest \cbk that ion heating is enhanced in two‑dimensional systems relative to one‑dimensional systems, owing to the increased efficiency of magnetic annihilation and reconnection within the current sheet.

In Section II, we present the basic structure of super‑critical parallel shocks based on PIC simulation results. One‑dimensional and two‑dimensional shock heating and particle acceleration are discussed in Sections II‑A and II‑B, respectively. Section III extends the case study of energy partitioning introduced in Section II to a wider range of shock parameters, and we examine the role of Alfvén waves and the associated magnetic reconnection. Section IV provides a summary and outlines future perspectives. Appendices A and B present linear plasma instability analyses for the ion‑cyclotron instability and the ion‑acoustic instability that appear in the shock region.
\section{Basic Structure of Parallel Shock Waves}
\begin{figure*}
\centering
\includegraphics[width=18cm]{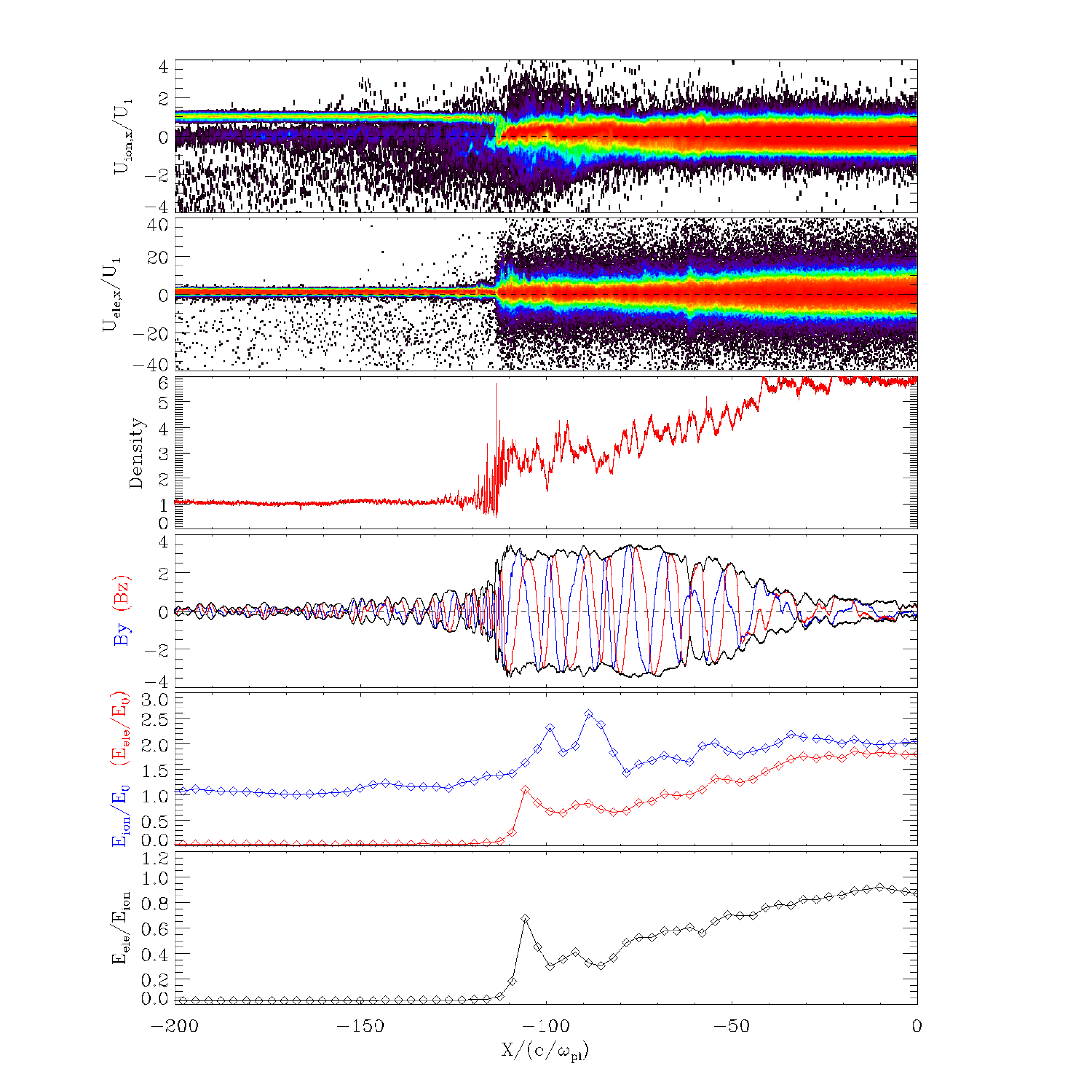}
\caption{Snapshot of a one-dimensional (1D) parallel shock with a non-relativistic upstream plasma flow of $u_0 = 0.4$. From top to bottom:  
(a) ion phase-space diagram in $(x, U_{\mathrm{ion},x})$;  
(b) electron phase-space diagram in $(x, U_{\mathrm{ele},x})$;  
(c) plasma density (red);  
(d) transverse magnetic fields $B_y$ (blue), $B_z$ (red), and their envelope $(B_y^2 + B_z^2)^{1/2}$ (black);  
(e) kinetic energies of ions (blue) and electrons (red); and  
(f) the ratio of ion to electron kinetic energies.  
The shock front is located near the center of the plot, with the upstream region on the left and the downstream region on the right. A portion of the simulation domain is shown, spanning from $X/(c/\omega_{\rm pi}) = -200$ in the upstream region to $X = 0$ at the downstream boundary.}
\label{fig:snap1d}
\end{figure*}
We investigate the energy partition between ions and electrons in collisionless parallel shock waves. A fraction of the upstream ions is reflected at the shock front by the shock potential and/or the magnetic mirror, and these reflected ions carry substantial energy that drives strong electrostatic and electromagnetic instabilities. Our analysis focuses on the large‑amplitude Alfvén waves generated by the reflected ions and on their further amplification as they traverse the shock transition. These waves efficiently scatter both ions and electrons, facilitating rapid conversion of bulk flow energy into thermal energy. Among the various plasma processes operating in high Mach number shocks, we place particular emphasis on plasma heating mediated by nonlinear wave dynamics, including parametric instabilities and magnetic reconnection within the current sheets formed by trains of large‑amplitude Alfvén waves \cite{Karimabadi14}.

It has been suggested that many current sheet and magnetic island structures can be formed in downstream and that magnetic reconnection/magnetic annihilation \crd may \cbk play an important role on plasma heating and particle acceleration \cite{Gingell23,Bessho23,Bessho25a,Bessho25b}. In our paper, plasma heating mechanisms are investigated by comparing results from one-dimensional (1D) and two-dimensional (2D) particle-in-cell (PIC) simulations. 

\subsection{Shock heating by large-amplitude Alfv\'en waves in 1D system}
Let us first study plasma heating in parallel shock waves by using one-dimensional PIC simulations.  As our main purpose in this paper is to understand a role of the current sheet on plasma heating, it is important to understand first the heating process under the condition without magnetic reconnection in the current sheets.

Shown in Figure \ref{fig:snap1d} is a snapshot of a parallel shock with the upstream four velocity $u_0=0.4$ and the magnetization parameter $\sigma=0.0158$.  Namely, the Alfv\'en Mach number $M_A=(u_0/\gamma_0)/\sqrt{\sigma/(\sigma+1)} \sim 3.06$ in the downstream plasma rest frame (The Alfv\'en Mach number in the shock rest frame is $M'_A \sim 3.86$).  Note that the Alfv\'en speed $V_A$ and $\sigma$ are, respectively, given by $V_A/c=\sqrt{\sigma/(\sigma+1)}$ and $\sigma=B_0^2/4 \pi n_0 (m_i+m_e) \gamma_0 c^2$, where $B_0$, $n_0$ and $\gamma_0$ are the magnetic field, the plasma density, and the Lorentz factor of plasma flow in upstream.  
In our simulations, a low-entropy, high-speed plasma composed of electrons and ions is injected from the left boundary and travels towards the positive $x$-direction. The downstream boundary at $x=0$ is modeled as a reflecting wall for both particles and waves. Through interactions between the incoming plasma and the reflected particles and waves from this boundary, a shock wave is generated and propagates backward in the negative $x$-direction.  Figure \ref{fig:snap1d} shows that the shock front is situated around the center of the plot $X/(c/\omega_{\rm pi}) =-110$ at time $t \Omega_{\rm ci}=134$. The leftward and rightward regions are the shock upstream and the shock downstream, respectively. In the simulation frame with this boundary setup, the average bulk plasma velocity in the downstream region becomes effectively zero. We focus on parallel shocks, where the upstream magnetic field is aligned with the plasma flow, namely, the initial uniform magnetic field $\vec{B}=(B_0, 0, 0)$. The mass ratio of ion to electron is $m_i/m_e=100$. We examine the plasma heating as the function of the upstream plasma flow and the magnetization parameter $\sigma$.
The average number of particles per cell in downstream and the ion inertia length $c/\omega_{\rm pi}$ are 512 particles and 230 grids, respectively.

From top to bottom in Figure \ref{fig:snap1d}, the ion phase-space diagram in $(X, U_{ix})$, the electron phase-space diagram in $(X, U_{ex})$, the transverse magnetic fields $B_y$ (red) and $B_z$ (blue), the plasma density, the kinetic energy of ions (blue) and electrons (red), and the ratio of the electron kinetic energy to the ion one.  $U_{ix}$ and $U_{ex}$ are normalized by $u_0$, and the magnetic field $B_y$ and $B_z$ are normalized by the upstream ambient magnetic field $B_0$.  The X-coordinate is normalized by the ion inertia length $c/\omega_{\rm pi}$.  The kinetic energy density $E_j$ is calculated in each bin summing the particle energies as $\sum_{\#} (\gamma_{\#,j}-1)m_j c^2$ for ions $j=i$ and electrons $j=e$, where $\gamma_{\#,j}$ is the Lorentz factor for each particle. They are normalized by the bulk kinetic energy of ion $E_0$ at the injection boundary.

The system size in the $X$-direction is not fixed; the left-hand boundary expands at the speed of light as time progresses. Figure \ref{fig:snap1d} displays only a portion of the simulation domain, spanning from $X/(c/\omega_{\rm pi}) =-200$ to $X=0$. From the top panel of the ion phase space diagram, we observe that a fraction of the incoming particles is reflected at the shock front, located around $X/(c/\omega_{\rm pi}) \sim -110$. The number density of the reflected ion is about $10\% \sim 30\%$ in the shock front.  The interaction between the low-entropy incoming ions and the reflected ions generates low-frequency, transverse whistler waves in the upstream region by the ion-beam cyclotron instability \cite{Gary81,Winske84,Hoshino85}.

As shown in the top and second panels, both ions and electrons are rapidly heated near the shock front located at $X/(c/\omega_{\rm pi}) \sim -110$. Further downstream, for $X /(c/\omega_{\rm pi}) > -60$, both species continue to be gradually heated in conjunction with plasma density compression (see the third panel from the top). The kinetic energies are presented in the second panel from the bottom: the blue and red curves represent ion and electron kinetic energies, respectively. In the region $-70 <X/(c/\omega_{\rm pi}) < -30$, where large-amplitude Alfvén waves gradually decay, electron heating becomes significant, whereas ion heating remains weak. As a result, the ratio of electron to ion kinetic energy density, shown in the bottom panel, approaches $E_{\rm ele}/E_{\rm ion} \sim 0.9$ in the downstream region.

\begin{figure*}
\centering
\includegraphics[width=16cm]{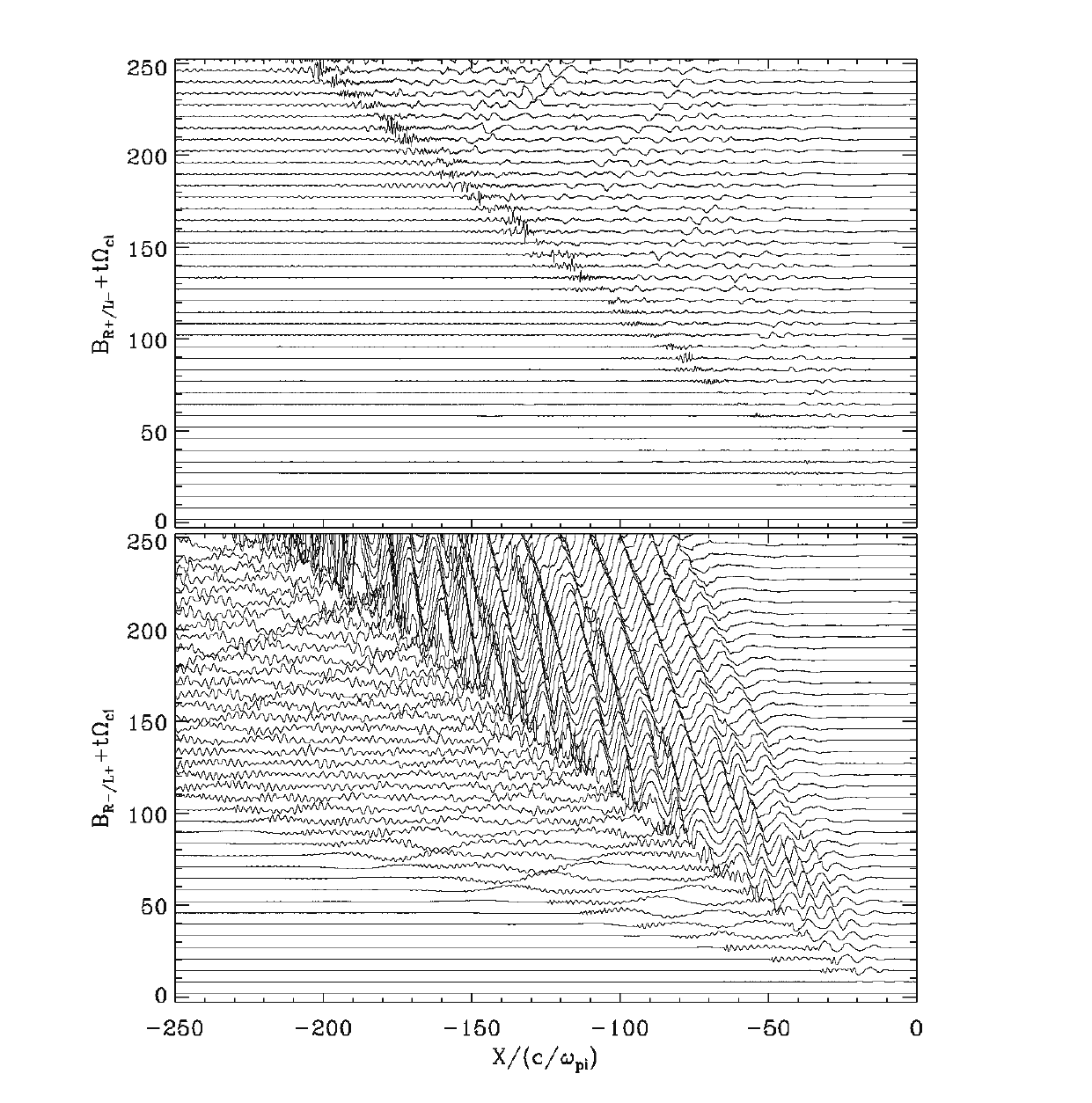}
\caption{Stack plot of Fourier-decomposed waveforms obtained from the transverse magnetic fields $B_y \pm i B_z$. The top and bottom panels represent $R^+/L^-$ (positive propagating, right-hand polarized mode/negative propagating, left-hand polarized mode) and $R^-/L^+$ (negative propagating, right-hand polarized mode/positive propagating, left-hand polarized mode), respectively.}
\label{fig:stack1d}
\end{figure*}

Roughly speaking, ion heating at the shock front results from the sudden change in bulk flow speed, and the associated heating energy is comparable to the upstream ion flow energy. In contrast, the energy gained by electrons after crossing the shock much exceeds the upstream electron bulk flow energy, because of the energy transfer from ion to electron through various plasma processes occurring in and around the shock front. One of them is the cross-shock potential induced by an inertia difference between ions and electrons, and it is known to convert about $10 \%$ of the ion kinetic energy into the electron heating \citep{Lembege04} by the cross-shock potential. The small-scale plasma waves and turbulence in the shock ramp are another important process for the electron heating \citep{Oka19,Amano22,Johlander23}.  In this paper, we focus on the gradual plasma heating associated with the large-amplitude Alfv\'en waves with several ten's ion inertia scales behind the shock front.

Shown in Figure \ref{fig:stack1d} is time evolution of the waveforms.  The horizontal axis represents both the time and the corresponding waveform, and the origin of the waveform is shifted in time as it is plotted.  The transverse waves are decomposed into right- and left-hand polarized modes by Fourier transformation for the original waves of $B_y \pm i B_z$. The top panel represents either $R^+$ (positive propagating, right-hand polarized mode) or $L^-$ (negative propagating, left-hand polarized mode), while the bottom panel shows either $R^-$ or $L^+$.  The horizontal axis is the $x$-coordinate normalized by the ion inertia length $c/\omega_{\rm pi}$, and the left- and the right-hand region corresponds to the upstream or downstream, respectively. The shock front is propagating from $(X/(c/\omega_{\rm pi}),B_{R/L}+t\Omega_{\rm ci})=(-30,35)$ to $(-210,250)$, and the shock propagation speed can be roughly estimated as $v_{sh} \sim 0.84 V_A \simeq 0.11 c$.

In upstream, we find that both the long wavelength wave and the short wavelength wave are generated in the bottom panel of $R^-/L^+$ mode, and it is clear that the long wavelength wave are propagating toward the shock front, namely, the wave mode is the $L^+$ mode.  The propagation speed can be estimated to be about $1.9 V_A \simeq 0.25 c$ from the slope of the wave node.

We think that both waves with the longer and shorter wavelengths can be generated by the ion-beam cyclotron instability \citep{Winske84,Hoshino85}.  Let us first assume the low-frequency approximation with $\omega \ll \Omega_{\rm ci}$.  The ion-beam cyclotron instability in the plasma rest frame can be simply understood by the resonance between the whistler wave $\omega =k V_A$ and the ion cyclotron motion with the super-Alfv\'enic speed with $\omega= k V_b -\Omega_{\rm ci}$, where $V_A$, $V_b$, and $\Omega_{\rm ci}$ are the Alfv\'en speed, the ion beam speed, and the ion cyclotron frequency, respectively. The ion beam energy is converted to the generation of the long-wavelength and low-frequency whistler wave (i.e., right-hand polarized MHD/Alfv\'en wave) with the wavenumber $k=\Omega_{\rm ci}/(V_b - V_A)$ and the frequency $\omega=V_A/(V_b- V_A) \Omega_{\rm ci}$. (The term of Alfv\'en wave is used to refer to low-frequency MHD wave regardless of its polarization).
However, if we take into account of the wave dispersion of the whistler wave of $\omega^2=(1+\omega/\Omega_{\rm ci})(k V_A)^2$ instead of $\omega = k V_A$, we may have another unstable whistler wave in a slightly higher frequency in addition to the above low-frequency whistler wave (see Appendix A).  

\begin{figure*}
\centering
\includegraphics[width=16cm]{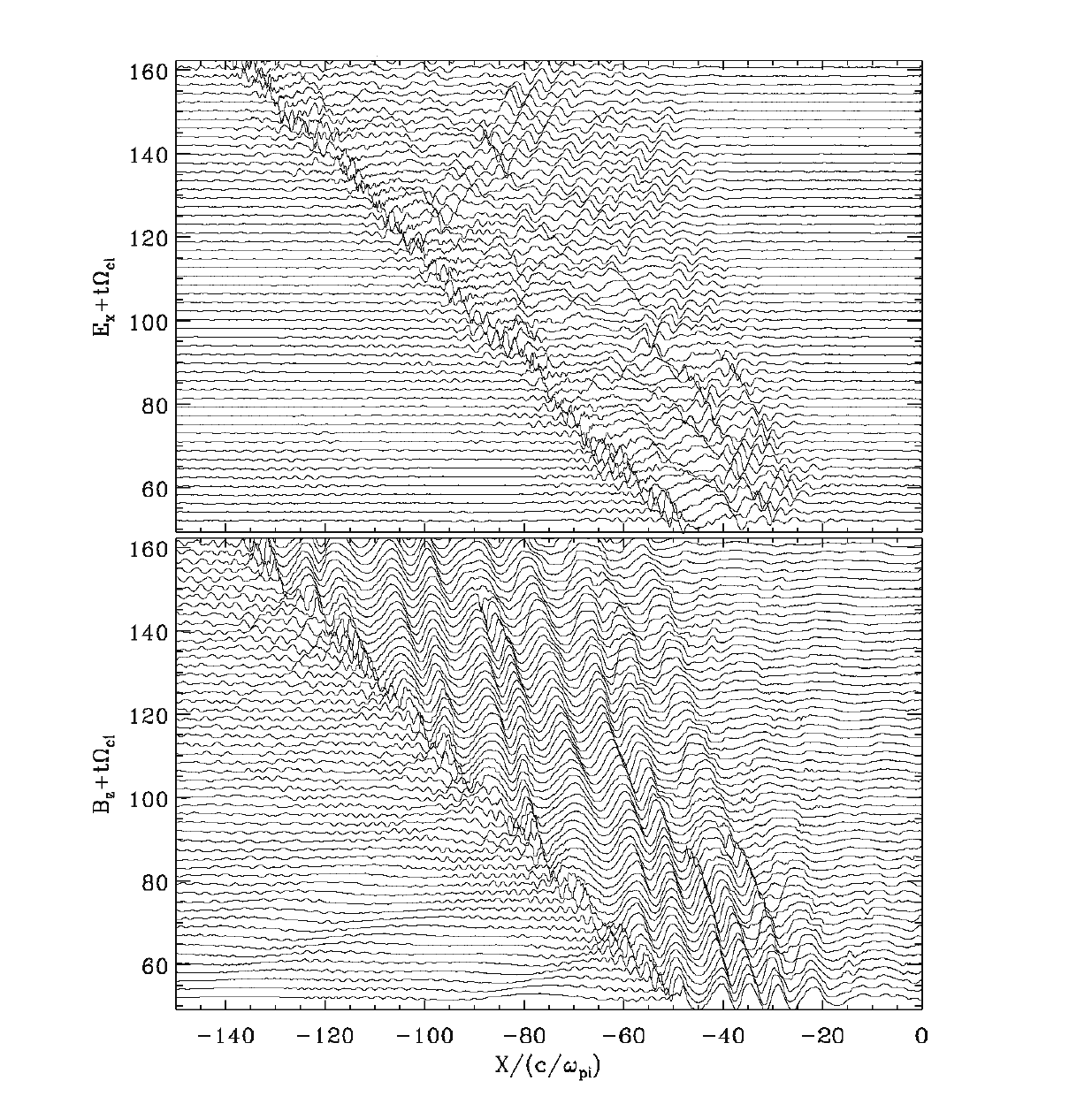}
\caption{Stack plot of waveform for the electric field $E_x$ (top) and the magnetic field $B_z$ (bottom).}
\label{fig:stack1d_ex}
\end{figure*}

Let us confirm whether the long wavelength wave is consistent with the above theory.
The long wavelength wave theoretically propagates negative $x-$direction with the speed of the Alfv\'en wave $V_A (= 0.13 c)$ in the upstream frame, but in the simulation frame the wave is convected by the upstream plasma flow speed of $v_0=u_0/\gamma_0=0.37c$, namely, the wave propagation speed becomes $(v_0-V_A)/(1-V_A v_0/c^2)=(0.37c -0.13c)/(1-0.13 \times 0.37) \sim 0.25 c \simeq 1.9 V_A$, which is close to the simulation result estimated from the slope in Figure \ref{fig:stack1d}.

Consequently, while the amplitudes of the transverse wave components $B_y/B_0$ and $B_z/B_0$ are less than unity in the upstream region, the transmitted waves after crossing the shock front are compressed together with the plasma density in downstream, and their amplitudes become several times larger than those in upstream.

The velocity of the higher frequency whistler waves may be faster than the upstream flow speed depending on the reflected beam speed from the shock front, and the higher frequency whistler waves may propagate toward the upstream direction. If the beam velocity is close to the Alfv\'en speed, the higher frequency waves are also convected to the downstream direction. As discussed Appendix A, the higher frequency whistler wave may be damped by the hot electrons escaping from the downstream. Therefore, the amplitude is not necessarily large, and the effect of its plasma heating may be weak.

In downstream in the right-hand side region, the transmitted waves across the shock front diverse into $R^+$ mode (top panel) and $R^-$ mode (bottom panel), and the amplitude of $R^-$ mode propagating upstream is larger than that of $R^+$ mode propagating downstream. This is consistent with the theoretical prediction \citep{McKenzie69}.

Downstream MHD waves may also be generated locally within the shock transition region when a velocity difference between the transmitted ions (the relatively cold component) and the hot downstream ions exists for high‑Mach‑number shocks with $M_A > 7$ \cite{Winske90,Krauss-Varban95,Scholer97,Burgess13}.
For lower or mildly high Mach numbers, however, the available free energy is insufficient to drive the ion‑cyclotron beam instability.  The downstream waves may be a mixture of waves generated both in upstream and in the transition region \cite{Scholer97,Burgess13}.

Under the circumstance of the large amplitudes of the transmitted whistler waves and the heated electrons at the shock front, the supra-thermal electrons are subject to the cyclotron resonance and the effective electron heating occurs through the condition of $\omega + k v_{ele}= \Omega_{ce}/\gamma_e$, where $v_{ele}$ and $\gamma_e$ are the supra-thermal electron velocity and its Lorentz factor, respectively.
As the Alfv\'en wave generated by the reflected ions has the frequency and wavenumber of $\omega=V_A/(V_b-V_A)\Omega_{\rm ci}$ and $k=\Omega_{\rm ci}/(V_b-V_A)$, the supra-thermal electron velocity required for resonance can be estimated as $v_{ele}\gamma_e \sim (V_b-V_A)(m_i/m_e)$. For our simulation parameters with $V_A/c \sim 10^{-1}$ and a mass ratio $m_i/m_e =100$, the resonance condition is satisfied for relativistic electrons with $v_{ele}\gamma_e > 10 c$. In realistic environments, however, where typically $V_A/c \sim 10^{-3}$, even mildly relativistic electrons with $v_{ele} \gamma_e \sim c$ can resonate with low-frequency whistler waves.

In addition, the large amplitudes of the transmitted $R^-$ waves are subject to modulational instability in a high-$\beta$ plasma. Figure \ref{fig:stack1d_ex} presents an expanded view of the time evolution of the waveforms; the electric field $E_x$ and the magnetic field $B_z$ are shown in the top and bottom panels, respectively. We observe that the Alfvén waves decay into other Alfvén waves with longer wavelengths. For example, while the dominant wavelength in the bottom panel is $\lambda _0 \sim 6 (c/\omega_{\rm pi})$ at $t\Omega_{\rm ci}\sim 70$, it evolves into $\lambda _1 \sim 10 (c/\omega _{pi})$ at $t\Omega _{ci}\sim 90$. Meanwhile, the compressional mode $E_x$ in the top panel exhibits a wavelength of $\lambda_2 \sim 15 (c/\omega_{\rm pi})$ at $t\Omega _{ci} \sim 70$. Roughly speaking, the corresponding wavenumbers are
$k_0=2\pi /\lambda _0 \sim 1.04 (\omega_{\rm pi}/c)$,
$k_1=2\pi /\lambda _1 \sim 0.63\, (\omega _{pi}/c)$, and
$k_2=2\pi /\lambda_2\sim 0.41\, (\omega _{pi}/c)$,
which satisfy the three‑wave coupling condition $k_0=k_1+k_2$.

\begin{figure*}
\centering
\includegraphics[width=16cm]{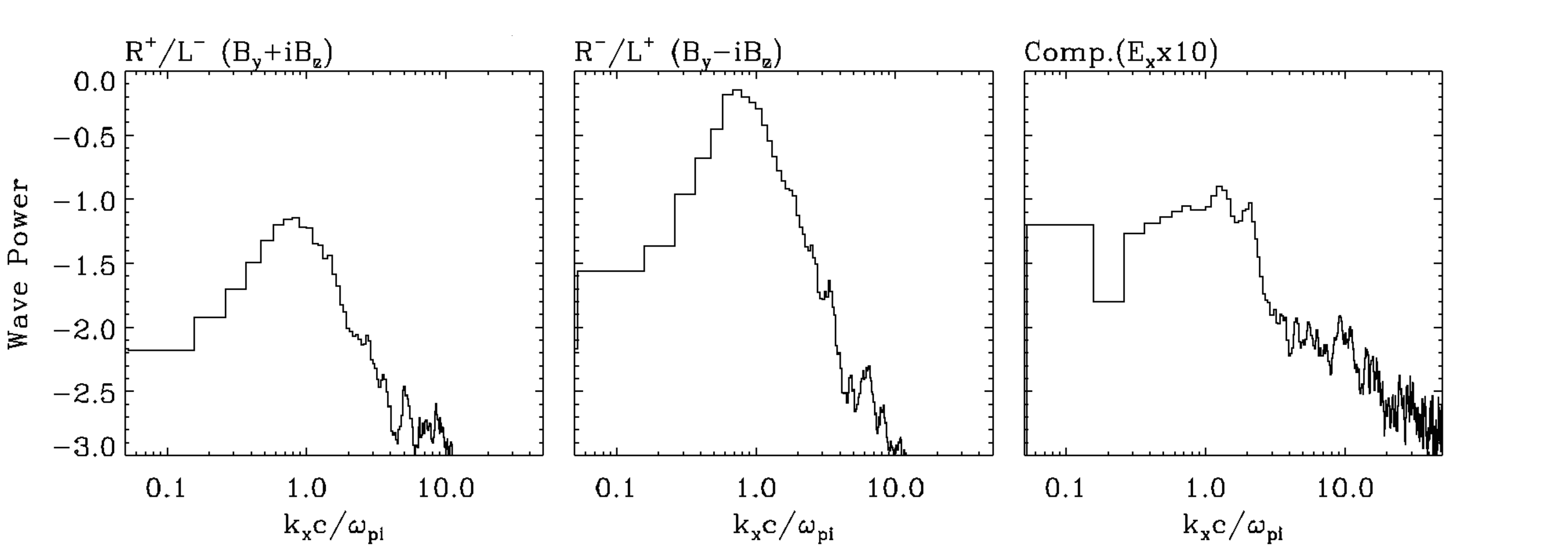}
\caption{Fourier wave power spectra of the transverse $R^+/L^-$ mode and $R^-/L^+$ mode, as well as the compressional mode $E_x$, in the downstream region at the same time as Figure~\ref{fig:snap1d}. Both the horizontal and vertical axes are shown on logarithmic scales.}
\label{fig:fft1d_hann}
\end{figure*}

Figure \ref{fig:fft1d_hann} shows the Fourier wave power spectra as a function of wavenumber $k_x c/\omega_{\rm pi}$ at the same time of $t\Omega _{\mathrm{ci}}=134$ as in Figure \ref{fig:snap1d}. Both the horizontal and vertical axes are shown on logarithmic scales.  The left, center, and right panels correspond to the transverse $R^+/L^-$ mode, the transverse $R^-/L^+$ mode, and the compressional mode of $E_x$, respectively.
The wave power of $R^+/L^-$ mode is obtained by taking Fourier transformation for $B_y + i B_Z$, while that of $R^-/L^+$ is calculated from $B_y - iB_z$.  Note that the wave power of $E_x(k)^2$ has been multiplied by a factor of 10 for clarity.
We select the spatial region behind the shock front from $X/(c/\omega _{\mathrm{ci}})=-110$ to $-50$ in order to avoid the abrupt change at the shock front. A Hann window is also applied to remove the boundary effect for both ends. The horizontal axis represents the wavenumber $k_xc/\omega _{\mathrm{ci}}$, and the vertical axis shows the power in logarithmic scale. In the spectrum of the $R^-$ mode (center panel), the peak power appears at $kc/\omega_{\mathrm{pi}}\sim 0.6$, and a comparable level of wave power extends up to $kc/\omega _{\mathrm{pi}}\sim 1$. On the other hand, the spectrum of the compressional wave $E_x$ exhibits a broadband distribution up to $kc/\omega _{\mathrm{pi}}\sim 2$, with a lower cutoff appearing around $kc/\omega _{\mathrm{pi}}\sim 0.2$.

The wave train observed in the downstream region $X/(c/\omega _{\mathrm{pi}})=-80\mathrm{\ to\ }-50$ around $t \Omega_{\rm{ci}} \sim 130$ in Figure \ref{fig:stack1d_ex} corresponds to the Fourier modes with $kc/\omega _{\mathrm{pi}}\sim 2$.  We may assume that the lower frequency part of $kc/\omega _{\mathrm{pi}} > 0.2$ appears to participate in the modulational instability, but the higher frequency part cannot be explained by the standard modulational instability. Instead, as briefly discussed in Appendix~B, ion acoustic waves as a compressional wave can be generated in the presence of bi‑streaming ions and a hot 
electron population \cite{Akimoto85}. From a theoretical point of view, the maximum growth rate of the ion–acoustic instability appears around $k v_{\rm df} / \omega_{\mathrm{pi}} \sim 1$, where $v_{\rm df}$ is the velocity difference between the two ion components. 
Roughly speaking, the waveform with $k c / \omega_{\mathrm{pi}} \sim 2$ observed in Figure~\ref{fig:stack1d_ex} is consistent with this theoretical expectation.

However, the compressional waves in Figure~\ref{fig:stack1d_ex} appear in the relatively deep downstream region, where the velocity difference between the two ion components is not necessarily large.
Up to this point, we have assumed that the two ion components follow Maxwellian distribution functions, although their actual velocity distributions are not necessarily Maxwellian. To better understand the generation of ion‑acoustic waves, a more detailed analysis that takes into account the multi-component ions and/or non‑Maxwellian behavior of the ions would be required.

In the above discussion, we briefly examined several processes that may contribute to the generation of Alfvén waves in the downstream region—namely, the ion‑cyclotron beam instability, the modulational instability of Alfvén waves, and the ion‑acoustic instability driven by bi‑streaming ions. These mechanisms provide possible explanations for the large‑amplitude compressional waves observed downstream, but they are not necessarily conclusive for fully understanding their origin. The downstream region is inherently turbulent and involves complex nonlinear coupling processes. Nevertheless, the modulational instability and the ion‑acoustic instability represent important processes that contribute to plasma heating.

As discussed in Appendix~B, a hot ion population can absorb the ion acoustic wave due to Landau damping.
It is also expected that the large amplitude, right-hand circularly polarized Alfv\'en wave can preferentially accelerate electrons in the course of successive parametric instabilities \cite{Matsukiyo09,Isayama23}.  In fact, while we observe the gradual increase of the electron kinetic energy $E_{\rm ele}$ from $X/(c/\omega_pi) \sim -80$ to $X/(c/\omega_{\rm pi}) \sim -30$ in Figure \ref{fig:snap1d}, the increase of the ion kinetic energy is limited.  As a result, the ratio of $E_{\rm ele}/E_{\rm ion}$ is gradually increases in association with the gradual decay of the Alfv\'en waves with $B_y$ and $B_z$.

\subsection{Shock heating including magnetic reconnection process in 2D system}
\begin{figure*}
\centering
\includegraphics[width=16cm]{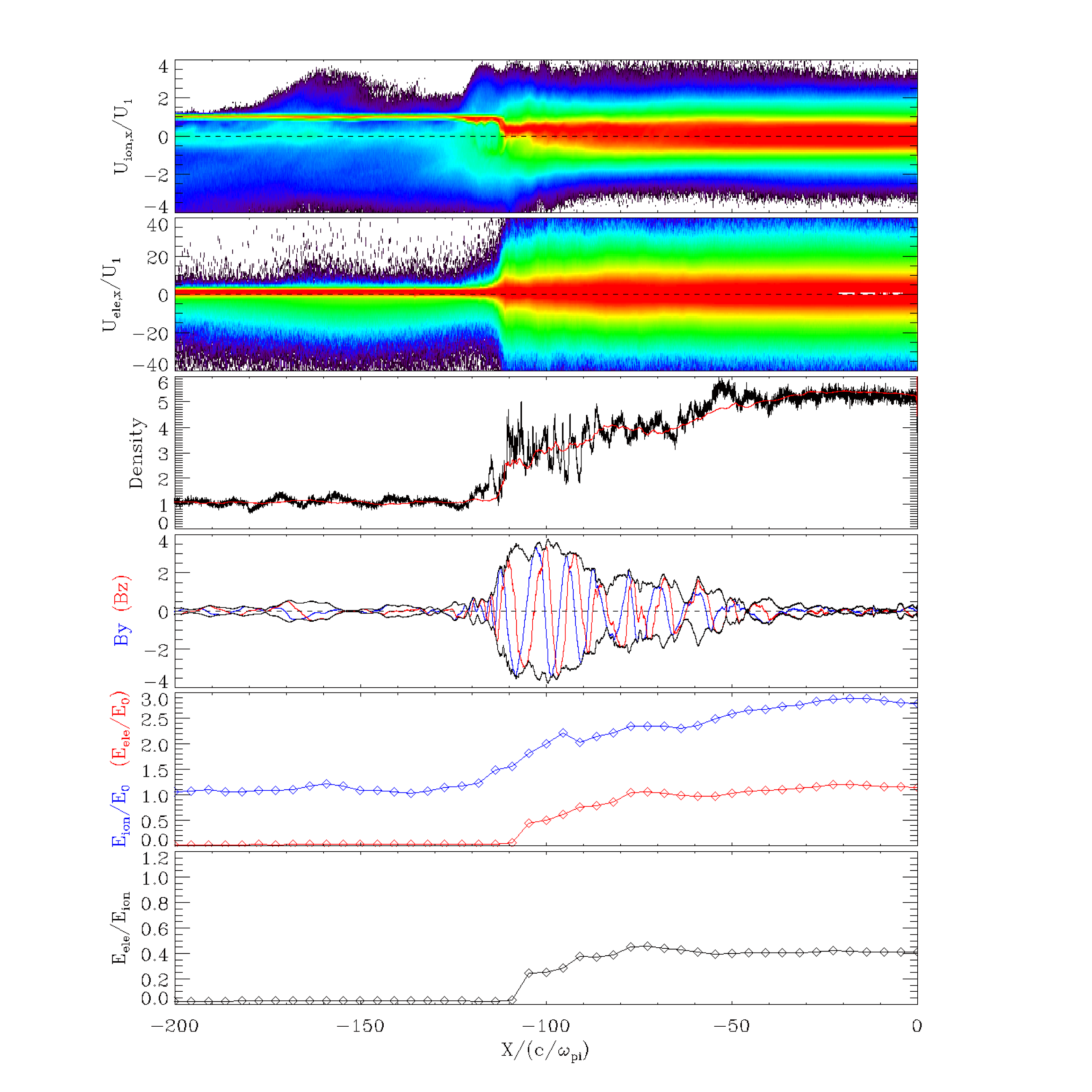}
\caption{Snapshot of a two-dimensional (2D) parallel shock obtained with the same physical parameters as in Figure~1. 
The system size in the $y$ direction is $26\,(c/\omega_{\rm pi})$.}
\label{fig:snap2d}
\end{figure*}

While we studied the electron energization during the decay of the Alfv\'en waves in the previous section, in this subsection we study the energy partitioning in two-dimensional system, where magnetic field energy dissipation by magnetic reconnection as 2D effects are expected to modify the energy transfer between ions and electrons. Figure \ref{fig:snap2d} is a snapshot of the parallel shock with the same physical parameters as shown in Figure \ref{fig:snap1d},
but the system size in the $y$, normalized in the upstream parameter, is $L_y = 26 c/\omega_{\rm pi}$ with the periodic boundary condition. (Because the downstream density is compressed, the effective downstream system size of $L_y$ becomes twice as large as that in the upstream region.) 
\cbk
The Alfv\'en Mach number is $M_A \sim 3.06$ in the downstream rest frame ($M'_A \sim 3.95$ in the shock rest frame). 
The average number of particles per cell in downstream and the ion inertia length $c/\omega_{\rm pi}$ are 64 particles and 230 grids, respectively. 

We observe the similar phase-space structures for both ions and electrons, and also observe the generation of the Alfv\'en waves in upstream and the large-amplitude compressed waves in downstream. The amplitude of the compressed waves in the 2D shock in Figure \ref{fig:snap2d} is almost the same as that observed in the 1D shock in Figure \ref{fig:snap1d}. The large-amplitude waves in Figure \ref{fig:snap2d}, however, quickly decay within about $20 c/\omega_{\rm pi}$, while the waves in Figure \ref{fig:snap1d} can keep the large amplitude more than $50 c/\omega_{\rm pi}$.

\begin{figure*}
\centering
\includegraphics[width=16cm]{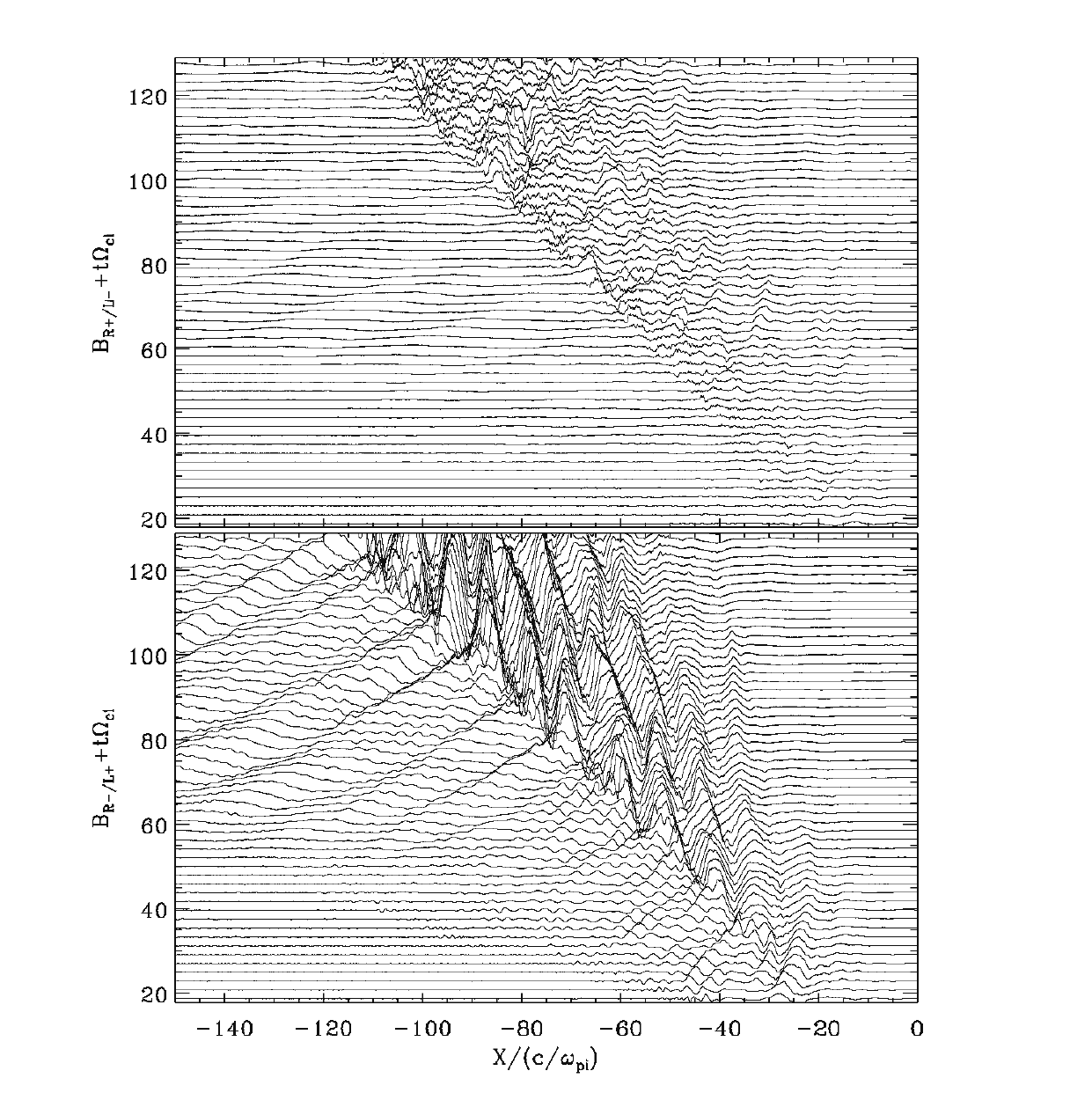}
\caption{Stack plot of Fourier-decomposed waveforms obtained from the transverse magnetic fields $B_y \pm i B_z$ at $y=0$.}
\label{fig:stack2d}
\end{figure*}

As we can see the dissipation of the Alfv\'en wave depends on the system size in $y$, and it is expected that the difference of the wave energy dissipation brings preferential ion and electron heating.  Two panels from the bottom in Figures \ref{fig:snap1d} and \ref{fig:snap2d} show the preferential heating in 1D and 2D shocks, respectively.  Figure \ref{fig:snap1d} with 1D shock shows $E_{\rm ion}/E_0 \sim 2.3$ and $E_{\rm ele}/E_0 \sim 1.6$, and then $E_{\rm ele}/E_{\rm ion} \sim 0.7$ in the downstream of $X/(c/\omega_{\rm ci}) = -40 \sim -20$ where the wave amplitudes become small, while Figure \ref{fig:snap2d} with 2D shock represents $E_{\rm ion}/E_0 \sim 2.8$ and $E_{\rm ele}/E_0 \sim 1.15$, and then $E_{\rm ele}/E_{\rm ion} \sim 0.4$ in the downstream of $X/(c/\omega_{\rm ci}) = -40 \sim -20$.

The case of 1D shock in Figure \ref{fig:snap1d} shows the electron kinetic energy $E_{\rm ele}$ increases inside the large-amplitude wave region, while the ion kinetic energy $E_{\rm ion}$ does not seem to increase. In addition, the gradual increase of the plasma density is associated with the large amplitude wave region. On the other hand, in the case of 2D shock in Figure \ref{fig:snap2d} both the ion and electron kinetic energies increase in association with the increase of the plasma density and with the decrease of the wave amplitudes.

In order to understand the wave property responsible to the ion and electron heating, Figure \ref{fig:stack2d} shows time evolution of the waveforms, which are decomposed into right- and left-hand polarized modes by Fourier analysis for the original waves of $B_y \pm i B_z$ with the same format as that shown in Figure \ref{fig:stack1d}. The top panel represents the $R^+$ or $L^-$, while the bottom panel shows $R^-$ or $L^+$.  The left- or right-hand region corresponds to the upstream or downstream, respectively, and the shock front is propagating from $(X/(c/\omega_{\rm pi}),B_{R/L}+t\Omega_{\rm ci})=(-30,35)$ to $(-120,140)$, the shock propagation speed is roughly $v_{sh} \sim 0.84 V_A \simeq 0.11c$.  We find that $L^+$ waves dominate in upstream in the simulation frame, and the propagation speed is about $1.67 V_A \simeq 0.22 c$ from the slope of the wave node.  This is close to the theoretical estimate of $0.25c$.  This result in 2D is almost same as that in 1D. 

In downstream, we can observe turbulent-like waves in the top panel, but the large-amplitude waves can be seen in the bottom panel of $B_{R^-/L^+}$ with the negative propagation, namely the waves with $R^-$ dominate. Therefore, hot electrons can easily satisfy the cyclotron resonance with the whistler waves of $R^-$, and can be further energized.  This is consistent with the property of the electron energization together with the decay of the transverse waves of $B_y$ and $B_z$ seen in Figure 1.  

\begin{figure*}
\centering
\includegraphics[width=16cm]{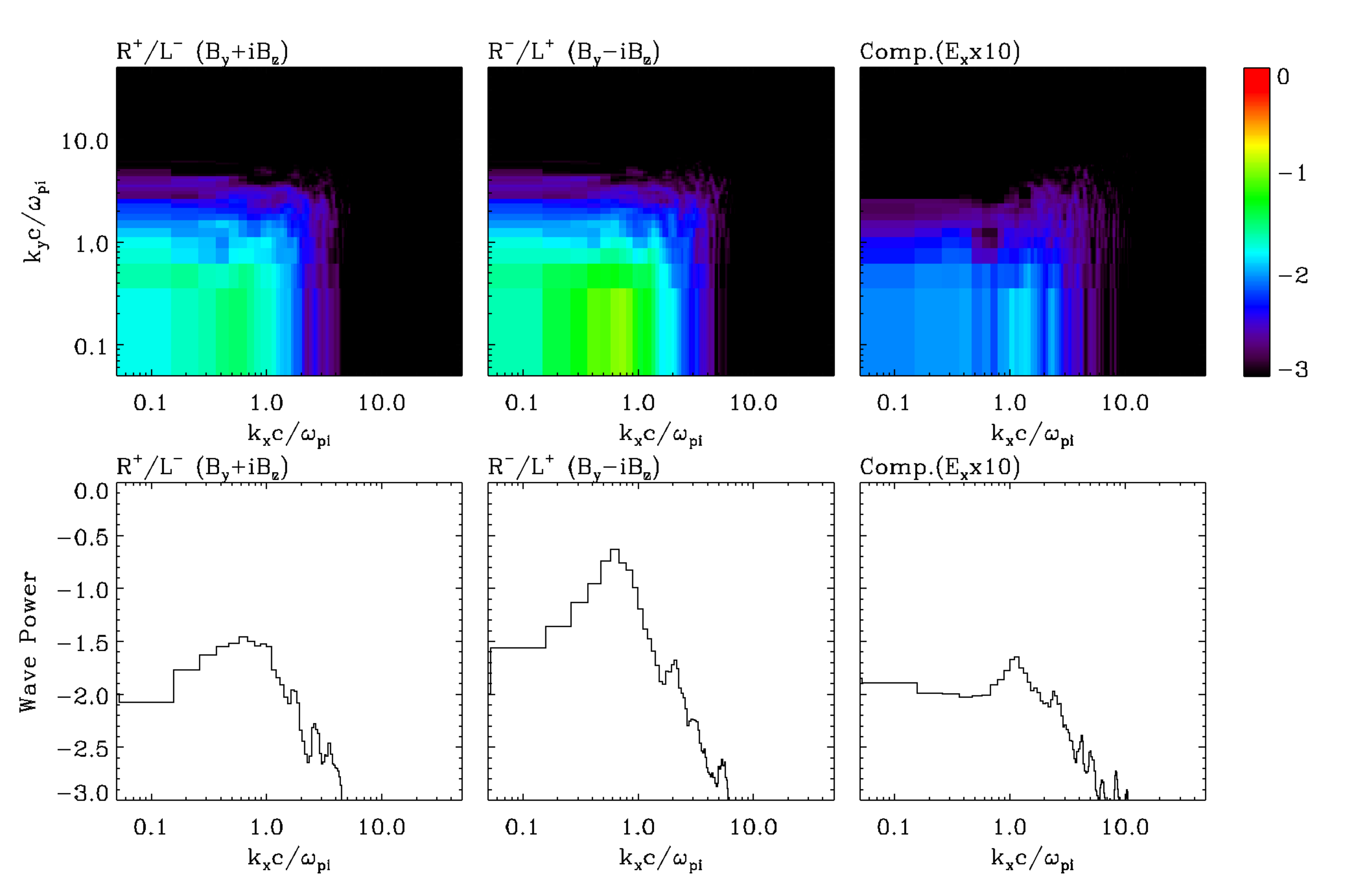}
\caption{Fourier wave power spectra at the same time as Figure~\ref{fig:snap2d}.
(Top) Two-dimensional spectra in $(k_x, k_y)$ space. (Bottom) One-dimensional slice along $k_y = 0$.
Both the wavenumbers and the wave powers are shown on logarithmic scales. }
\label{fig:fft2d_hann}
\end{figure*}
In order to clarify the similarity and differences between the 1D and 2D shocks, we present the Fourier spectra for two-dimensional shock in Figure \ref{fig:fft2d_hann}. The top three panels show the wave spectra in two-dimensional wavenumber space of $(k_x,k_y)$. The right-hand color bar shows the wave power in logarithmic scale.
We select the spatial region behind the shock front from $X/(c/\omega _{\mathrm{pi}}) =-110$ to $-50$, which is same as the wave spectra for 1D shock, and the Hann windows was applied. The bottom three panels are the one-dimensional slice at $k_y=0$.  The format is same as Figure \ref{fig:fft1d_hann}.  Note that the wave power of $E_x(k)^2$ has been multiplied by a factor of 10 for both 1D and 2D plots.
From both the 1D and 2D Fourier spectra, together with Figures \ref{fig:stack1d} and \ref{fig:stack2d},
we confirm that the dominant Alfvén waves are the $R^-$ mode with wavenumber $k_x c/\omega _{\mathrm{pi}} \sim 0.7$. In the right-hand panels for both 1D and 2D cases, we find that long-wavelength compressional modes are excited.  It is difficult to obtain the wavenumber of the compressional mode, but we can say that the wavenumber is around $k c/\omega_{\rm pi} = 0.2 \sim 0.4$.  These features are consistent with the interpretation that the modulation instability leads to the decay of Alfvén waves in the downstream region.

\begin{figure*}
\begin{center}
\includegraphics[width=16cm]{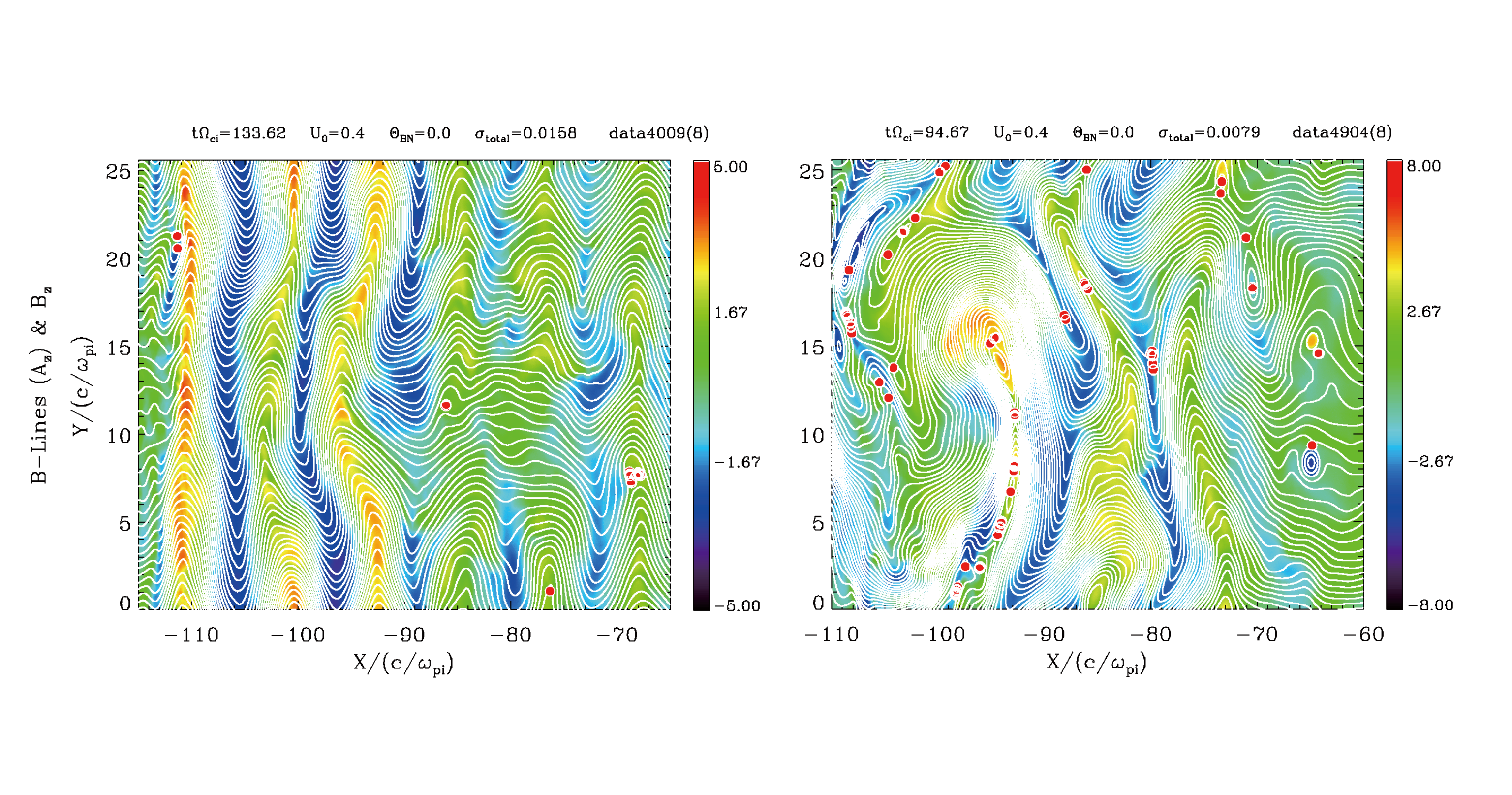}
\caption{Snapshot of the magnetic-field structure behind the shock front. The white lines show magnetic field lines in the $x$–$y$ plane, and the color contour represents the amplitude of $B_z/B_0$. The left and right panels correspond to the 
cases with $\sigma = 0.0158$ and $\sigma = 0.0079$, respectively.}
\label{fig:sheet}
\end{center}
\end{figure*}

We observe almost similar properties along $x-$structure, but we find that the wave power extends in the $y$ direction for 2D shock as we expected. It is therefore important to analyze the two-dimensional magnetic field structure. Figure \ref{fig:sheet} shows the magnetic field lines in behind the shock front region, plotted as contours of the vector potential $A_z(x,y)$ (white lines), together with the magnitude of $B_z/B_0$ in color. The left and right panels are the case with $\sigma=0.0158$ with $M_A \sim 3.06$ and $0.0079$ with $M_A \sim 4.19$, respectively.
The left panel is the same simulation result as that in Figure \ref{fig:snap2d}.
The red points indicate the local saddle points of the vector potential $A_z(x,y)$, \crd which may correspond to \cbk the X-point of magnetic reconnection.

By comparing two cases with $\sigma=0.0158$ in the left panel and $\sigma=0.0079$ in the right panel, we find that many current sheets are formed and become more turbulent for the lower magnetization parameter for $\sigma=0.0079$. A large number of the X-point with the local saddle point can be seen as well.
On the other hand, for the case with $\sigma=0.0158$ in the left panel, the saddle points (i.e., X-point of \crd the magnetic field configuration \cbk) can be seen only around $(X/(c/\omega_{\rm pi}), Y/(c/\omega_{\rm pi})=(-112,21)$, $(-85,11)$ and $(-70,8)$, but we observed the positions of X-point changes with time and the \crd X-point regions \cbk occur sporadically in a wide downstream region (not shown here). Therefore, the sparse or dense magnetic field lines may suggest that the tearing mode with the normal magnetic field \cite{Galeev76,Coroniti80,Buchner89,Zelenyi08,Pritchett10} \crd may occur \cbk and is dissipating the magnetic field energy in a wide downstream region.  

\begin{figure*}
\centering
\includegraphics[width=16cm]{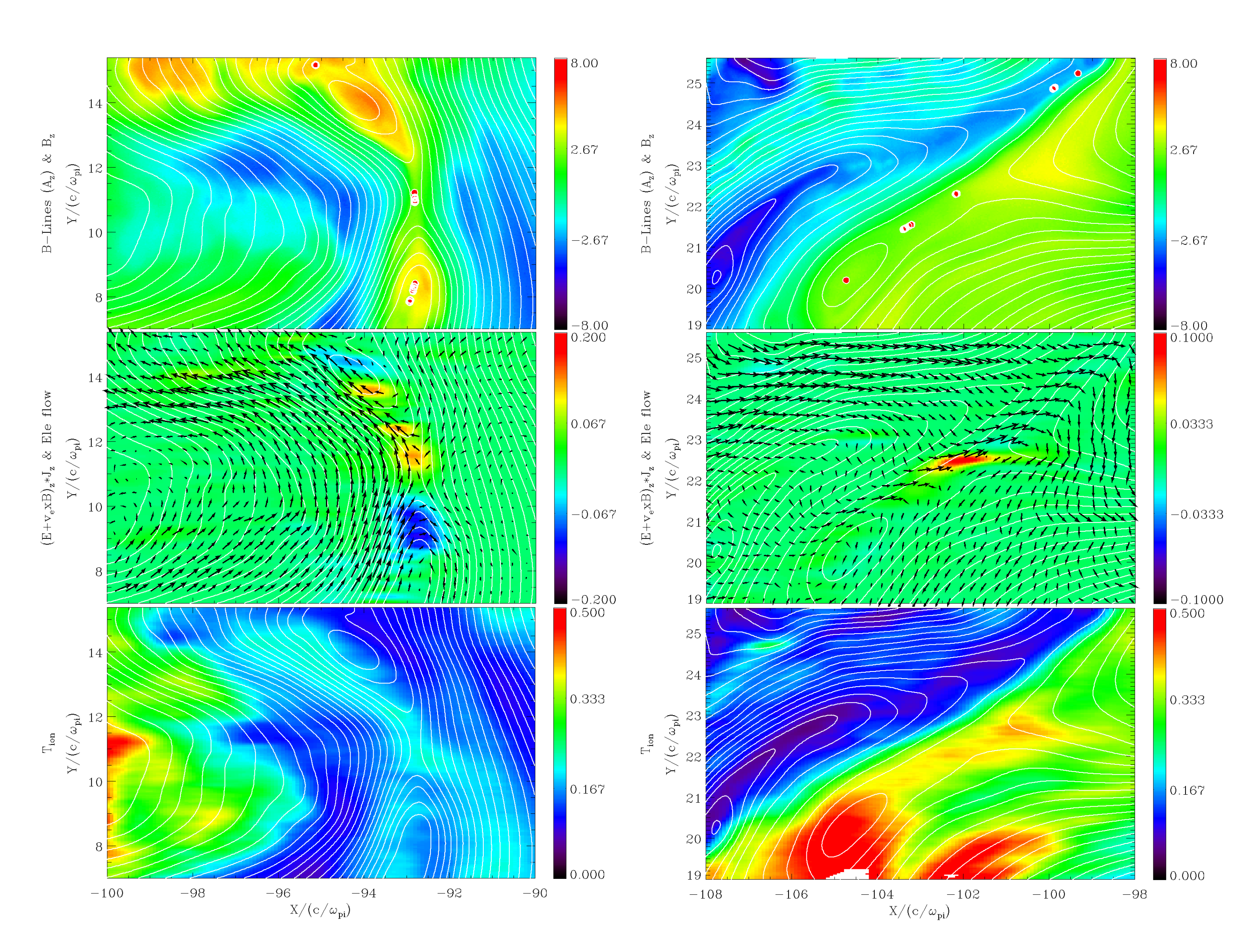}
\caption{Snapshot of the magnetic-field structure (top), the Ohmic dissipation (middle), and the ion temperature (bottom). In the top panel, the white lines show magnetic field lines in the $x$–$y$ plane, and the color contour represents the amplitude of $B_z/B_0$. In the middle panel, the black arrows show the electron flow pattern in the $x$-$y$ plane and the color contour represents the Ohmic dissipation of $J_z \cdot E'_z$.  \crd In the bottom panel, the ion temperature is shown by the color contour together with the magnetic field lines\cbk.  The left and right panels correspond to two \crd X-type \cbk regions around $(X,Y)/(c/\omega_{\rm pi}) \sim (-93,11)$ and $(-102,22.5)$ in Figure~\ref{fig:sheet}, respectively.}
\label{fig:sheet2}
\end{figure*}

We examine the Ohmic dissipation around the X-point region using $\vec{J} \cdot \vec{E}'$, where $\vec{E}'= \vec{E} + \vec{v}_e \times \vec{B}/c$ and $\vec{v}_e$ denote the electric field in the electron frame and the electron velocity. This quantity is widely used as a diagnostic for identifying the active magnetic diffusion region through the violation of the electron frozen-in condition \cite{Vasyliunas75,Hesse99,Zenitani11}. 
Figure~\ref{fig:sheet2} shows enlarged views of two X-type regions for the case $\sigma = 0.0079$ presented in Figure~\ref{fig:sheet}. The top and bottom panels display the magnetic-field structure and the Ohmic dissipation estimated by $J_z \cdot E'_z$, respectively. The left and right columns correspond to the reconnection sites located at $(X,Y)/(c/\omega_{\rm pi}) \approx (-93, 11)$ and $(-102, 22.5)$, respectively. The black arrows in the bottom panels indicate the electron flow vectors in the $x$-$y$ plane, and the color contour is the value of $J_z \cdot E'_z$, normalized by $e n_0 c B_0$ where $n_0$ and $B_0$ are the upstream plasma density and the upstream magnetic field, respectively.

Our \crd X-type regions, which may correspond to the magnetic reconnection, \cbk are not necessarily standard isolated reconnection sites, and the inflow and outflow structures are influenced by other plasma processes occurring in the shock downstream. Nevertheless, as seen in the two panels, strong Ohmic dissipation with positive $J_z \cdot E'_z$ values arises in and around the X-type region. The flow pattern around the X-point is also similar to that of guide-field magnetic reconnection \cite{Pritchett04}.
Although the analyses of the plasma flow around the X-type magnetic field structure and $J_z \cdot E'_z$ signature are only circumstantial evidence of magnetic reconnection, and although other diagnostics based on the velocity distribution function such as non-gyrotropic/agyrotropic distribution \cite{Hoshino98,Scudder12} can also be used to identify the diffusion region,
\cbk
our analyses suggest that reconnection constitutes one of the important energy conversion processes in the shock downstream.

\crd
Furthermore, in the bottom panel, color contours of the ion temperature normalized by the rest mass energy are shown.  Since the ion distribution functions exhibit non-gyrotropic and anisotropic features, the ion temperature $T_{ion}$ is evaluated as the average of three diagonal components of the temperature tensor, i.e., $T_{ion}=(T_{xx}+T_{yy}+T_{zz})/3$.  We find that the hot ions are generated around two reconnection regions located at $(X,Y)/(c/\omega_{\rm pi}) \sim (-93,11)$ and $(-102,22.5)$. These heated ions are subsequently ejected toward the corresponding O-type regions.  The ion temperature inside the plasma sheet is higher than that outside the plasma sheet. The hot plasma jet extending from $(X,Y)/(c/\omega_{\rm pi}) \sim (-93,11)$ to $(-95,14)$ is slightly shifted to the left of the O-type region. The reconnection jet extending from $(X,Y)/(c/\omega_{\rm pi}) \sim (-102,22.5)$ to $(-100,23)$ ends before reaching the center of the corresponding O-type region.  Because highly complex processes are occurring around the reconnection regions, the ion heating patterns are not necessarily same as that expected for standard isolated reconnection. Nevertheless these results provide further evidence that magnetic reconnection in the shock downstream contributes to ion heating.
\cbk

We note that some regions exhibiting an X-type magnetic field configuration do not necessarily show positive energy dissipation in $J_z \cdot E'_z$ and \crd hot plasma production\cbk. For example, the X-type region around $(X,Y)/(c/\omega_{\rm pi}) \sim (-100, 25)$ in the right panel of Figure~\ref{fig:sheet2} does not exhibit positive $J_z \cdot E'_z$ and \crd hot plasma production\cbk. This indicates that reconnection regions with an X-type magnetic field geometry are strongly influenced by the complex interplay of turbulence, wave activity, and local plasma conditions. The coalescence of magnetic islands is also one of the important processes contributing to the turbulence in the downstream region \cite{Hoshino12}. Our interpretation is therefore that reconnection is present and plays a meaningful role in the downstream dynamics, but it does not account for all observed differences between the 1D and 2D simulations. 

The Alfvénic waves generated upstream are compressed across the shock front, and the compressed magnetic fields form current sheets with a turbulent magnetic-field structure. The formation of such current sheets has been discussed by several authors \cite{Karimabadi14,Bessho23,Bessho25a,Bessho25b}. In particular, Bessho et al. \cite{Bessho23,Bessho25a,Bessho25b} demonstrated that magnetic reconnection at both ion and electron scales contributes to plasma heating and particle acceleration in quasi-parallel shocks.  They discussed the detailed heating and acceleration processes in association of the reconnection jets by paying attention to the kinetic reconnection behavior.  It has also been suggested that the efficiency of particle acceleration can be enhanced in the multi-current-sheet systems where numerous reconnection events occur \cite{Hoshino12}.

Focusing solely on the energy partitioning between ions and electrons during magnetic reconnection, it is well established that in non‑relativistic reconnection, where the Alfvén speed is much smaller than the speed of light, most of the released magnetic energy is converted into ion heating. For large‑scale reconnection with a well‑developed reconnection jet, ions and electrons share the same bulk velocity in the Alfvénic plasma outflows, so the bulk kinetic energy is predominantly carried by the ions. In addition, ions are preferentially heated in the magnetic diffusion region through non‑adiabatic Ohmic heating \cite{Hoshino18}. Such preferential ion heating has been observed in the Earth's magnetotail \cite{Eastwood10,Imada11} and in simulation studies \cite{Drake12,Haggerty15,Shay18,Hoshino18,Liu25}.
In contrast, in relativistic reconnection, both ions and electrons can be efficiently heated in and around the magnetic diffusion region, and well‑developed nonthermal particle populations are commonly produced \cite{Hoshino22,Hoshino23,Guo24,Totorica24}. Considering these established properties of magnetic reconnection, the preferential ion heating observed in the downstream region—associated with the decay of large‑amplitude Alfvén‑wave structures—can be interpreted as a natural consequence of energy dissipation via non‑relativistic reconnection.

From our case study with upstream flow speed $u_0=0.4$ and Alfvén Mach number $M_A=3.04$, we find that among the various plasma heating processes involved in high–Mach-number shocks, the major difference between two-dimensional and one-dimensional shocks lies in the ion heating \crd probably \cbk produced by current sheets and their magnetic reconnection/magnetic annihilation. As a result, the downstream ion-to-electron
``effective'' \cbk
temperature ratio becomes significantly larger in two-dimensional shocks than in one-dimensional shocks.

\begin{figure*}
\centering
\includegraphics[width=10cm]{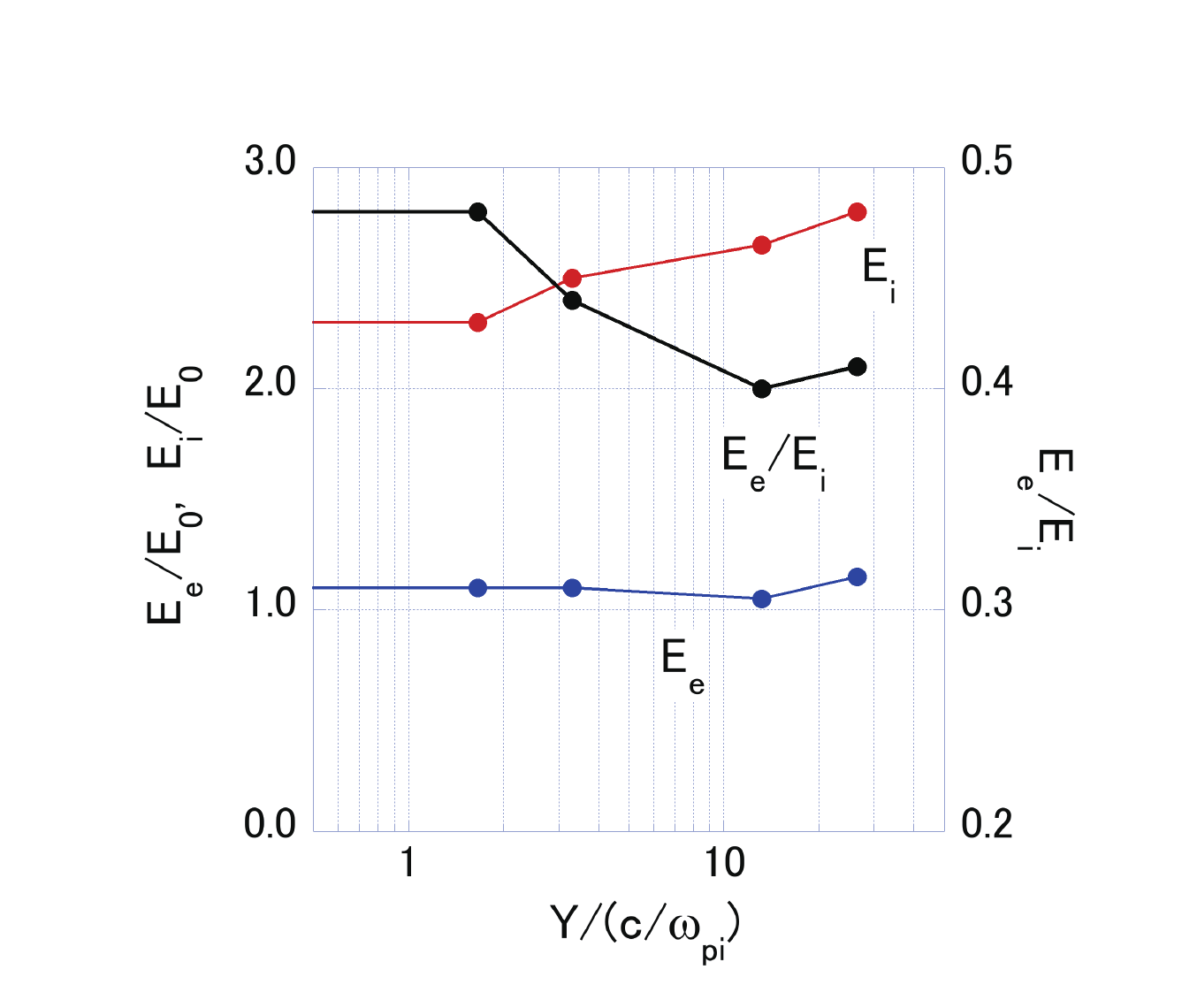}
\caption{Dependence of the downstream kinetic energies of ions ($E_i$, red) and electrons ($E_e$, blue), as well as the electron-to-ion kinetic energy ratio $E_e/E_i$ (black), on the system size $Y/(c/\omega_{\rm pi})$ for an upstream flow velocity of $u_0 = 0.4$ and magnetization parameter $\sigma = 0.0158$. 
The kinetic energies $E_i$ and $E_e$ are normalized by the upstream kinetic energy $E_0$.}
\label{fig:survey2}
\end{figure*}

It is interesting to examine how the energy partitioning is affected by the size of the $y$ coordinate. Figure~\ref{fig:survey2} shows the dependence of the ion and electron kinetic energies, as well as the ratio of electron to ion kinetic energy in the downstream region, as a function of the system size $Y/(c/\omega_{\rm pi})$.  The kinetic energies obtained in the one-dimensional result are almost identical to those for the case $Y/(c/\omega_{\rm pi}) = 1.57$, and the corresponding horizontal reference lines are drawn below $Y/(c/\omega_{\rm pi}) = 1.57$. 
While the electron kinetic energy does not exhibit a strong dependence on the $y$ size, the ion kinetic energy gradually increases for $Y/(c/\omega_{\rm pi}) > 1.57$. 
The ratio of electron to ion kinetic energy, $E_e/E_i$, appears to saturate around $0.4$.

In order to capture reconnection phenomena in which the thickness of the reconnecting current sheet becomes comparable to the ion inertial length near the X-point, the system size must exceed the current-sheet thickness by at least an order of magnitude. The above results are consistent with the standard reconnection picture.

\section{Parameter Survey}
In the case study discussed in the previous section, we showed that the energy partitioning between ions and electrons differs significantly between one-dimensional and two-dimensional shocks. In particular, ions can be strongly energized in two-dimensional shocks through magnetic energy dissipation in the current sheet formed by a train of large-amplitude Alfvén waves. It is therefore important to examine whether such preferential ion heating occurs over a wide range of parameters. The generation of large-amplitude Alfvén waves is a ubiquitous phenomenon \cite{Kennel85,Paschmann81,Leroy81}, and their amplitude depends on the number density of the reflected ions and the magnetization parameter $\sigma$ or the Alfv\'en Mach number $M_A$. The relative importance of plasma heating via reconnection may also depend on $\sigma$.  In this section, we investigate the parameter dependence of the energy partitioning.

\begin{figure*}
\centering
\includegraphics[width=16cm]{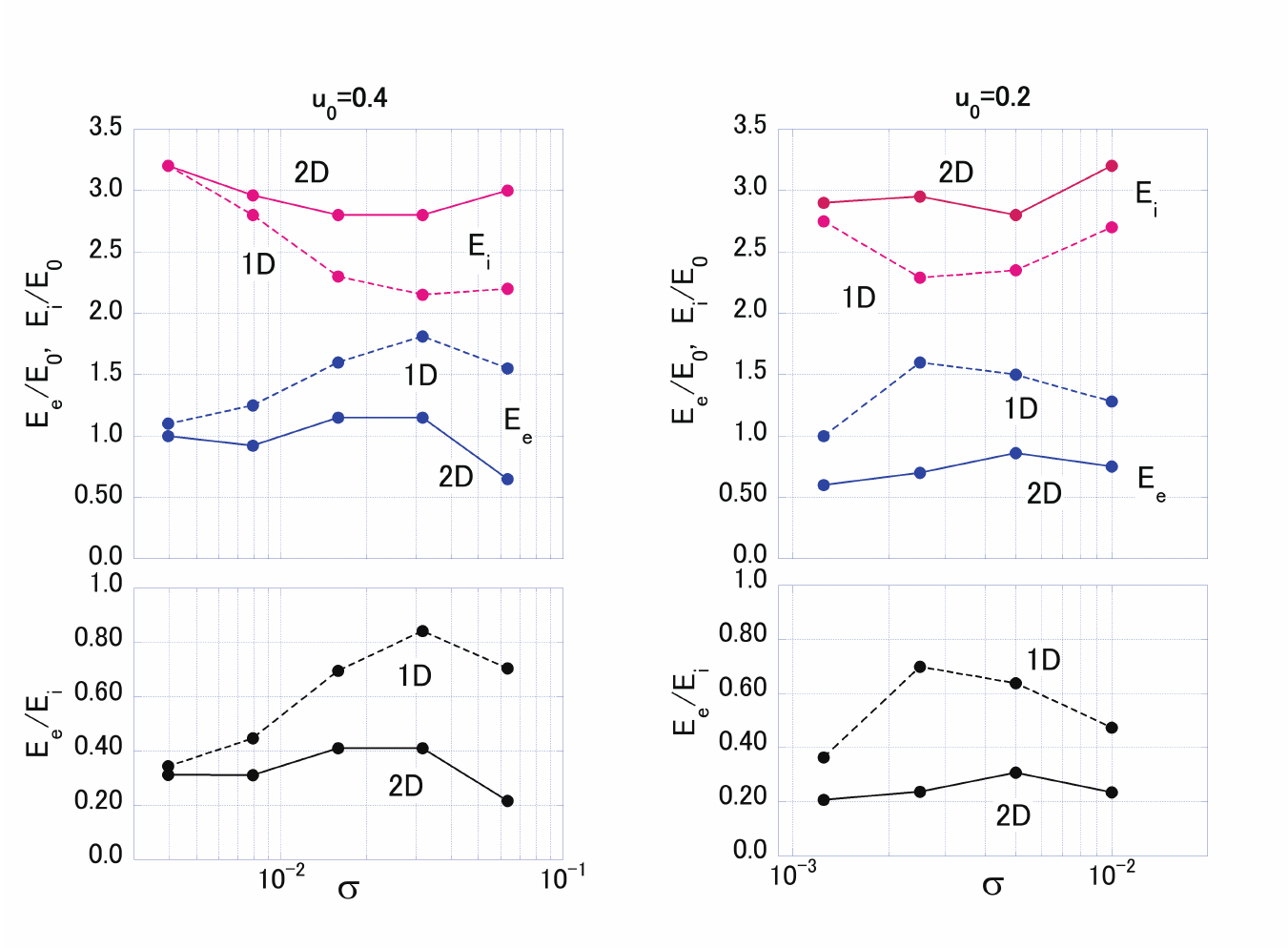}
\caption{Dependence of the ion and electron kinetic energies on the magnetization parameter $\sigma$. 
The left-hand panels show $E_i/E_0$ and $E_e/E_0$ (top) and the ratio of electron to ion kinetic energies $E_e/E_i$ 
(bottom) for an upstream flow velocity of $u_0 = 0.4$, while the right-hand panels show the corresponding results for 
$u_0 = 0.2$.}
\label{fig:survey}
\end{figure*}

Figure \ref{fig:survey} shows the downstream kinetic energies of ions, $E_i$, and electrons, $E_e$, as functions of the magnetization parameter $\sigma$. The solid and dashed lines represent the 2D and 1D results, respectively. The left and right panels correspond to upstream flow velocities of $u_0=0.4$ and $u_0=0.2$, respectively.

For the case of $u_0=0.4$ in the left panel, $\sigma =4 \times 10^{-3}$ corresponds to an Alfvén Mach number $M_A=v_0/V_A \sim 5.9$ where $v_0=u_0/\sqrt{1+u_0^2}$ and $V_A=\sqrt{\sigma/(1+\sigma)}$. (In the shock rest frame, $M'_A \sim 7.82$ and $7.06$ for one-dimensional and two-dimensional shock, respectively.)  The case of $\sigma =6.4 \times 10^{-2}$ and $u_0=0.4$ corresponds to $M_A \sim 1.5$ in the downstream rest frame. (In the shock rest frame, $M'_A \sim 2.00$ for 1D shock and $2.18$ for 2D shock).

For the case of $u_0=0.2$ in the right panel, $\sigma =1.25 \times 10^{-3}$ gives $M_A \sim 5.5$ ($M'_A \sim 7.01$ for 1D shock, and $M'_A \sim 6.39$ for 2D shock), and $\sigma =10^{-2}$ gives $M_A \sim 2.0$ $M'_A \sim 2.75$ for 1D shock, and $M'_A \sim 2.64$ for 2D shock).

We find that the ion kinetic energy exceeds the electron kinetic energy over a wide range of $\sigma$ for both $u_0=0.2$ and $0.4$. The ratio $E_e/E_i$, i.e., the energy partitioning between electrons and ions for small $\sigma$, however, is almost same between the 1D and 2D simulations.  For example, $E_e/E_i \sim 0.3$ for $u_0=0.4$ and $\sigma=4 \times 10^{-3}$, and $E_e/E_i \sim 0.2 - 0.35$ for $u_0=0.2$ and $\sigma=1.25 \times 10^{-3}$.
For those small $\sigma$, i.e., $\sigma < 4 \times 10^{-3}$ for $u_0=0.4$ and $\sigma < 10^{-3}$ for $u_0=0.2$, Alfvén waves with larger amplitudes normalized by the ambient magnetic field $B_0$ are generated, and strongly turbulent magnetic fields are also observed. However, the energy density of the Alfvén waves in low-$\sigma$ shocks is not large compared with the bulk flow energy density, and then the contribution of ion heating from turbulent current sheets generated by compressed Alfvén waves in the downstream region is not significant as well. Consequently, the energy partitioning between ions and electrons does not differ between the 1D and 2D cases for the small $\sigma$. 

For relatively larger $\sigma$ shocks, where the Alfvén Mach number is not necessarily high, the turbulence level becomes moderate compared with higher Mach number shocks, and the current sheets formed by compressed Alfvén waves are not necessarily strong. As shown in Figure \ref{fig:sheet}, the current sheet for $\sigma =0.016$ in the left panel exhibits a dipole-like structure of the anti-parallel magnetic field structure, whereas that in the right panel shows an elongated structure. This suggests that reconnection activity is reduced for larger $\sigma$.
It is known that for a dipole-like current sheet, namely for a quasi-neutral current sheet with a finite normal magnetic component, the adiabatic electrons magnetized in the normal magnetic field may stabilize the tearing mode \cite{Galeev76,Coroniti80,Buchner89,Zelenyi08,Pritchett10}.   In our parameter survey of the super-critical Mach number shock with the reflected ions, the current sheet formed by the compressed Alfv\'en wave still provides more ion heating in two-dimensional system.

\section{Summary and Discussions}
In collisionless shock waves, the ion and electron temperatures are not necessarily equal, and in high–Mach-number shocks various plasma heating mechanisms operate on ions and electrons. The physical processes that govern their energy partitioning, however, remain elusive. In this paper, we aim to clarify this issue for parallel shock waves by focusing on the energy dissipation associated with magnetic reconnection in current sheets generated by a train of Alfvén waves.  \crd We suggest that ion heating due to magnetic reconnection is one of the important processes controlling the energy partitioning between ions and electrons\cbk.

In supercritical shocks \cite{Kennel85}, where incoming ions can be reflected back into the upstream region, large-amplitude Alfvén waves are generated by the ion-beam cyclotron instability \cite{Gary81,Winske84}. These waves are transmitted into the downstream region and further amplified by plasma compression. Ions and electrons can be heated by the amplified Alfvén waves through the modulation instability, and electrons can absorb the associated high-frequency whistler waves via electron cyclotron damping (see Appendix A).

In multi-dimensional systems, however, magnetic reconnection \crd may \cbk play an important role in dissipating the amplified magnetic fields in the shock downstream. During the compression of large-amplitude Alfvén waves, numerous current sheets with anti-parallel magnetic-field configurations can be formed.  \crd During the magnetic field energy dissipation, we suggested that magnetic reconnection contribute to ion plasma heating.  In fact\cbk, ions are known to be preferentially heated in and around the magnetic diffusion region in non-relativistic plasmas \cite{Eastwood10,Imada11,Haggerty15,Hoshino18}.  In well-developed reconnection, Alfvénic ion jets can carry most of the kinetic energy, and reconnection is expected to preferentially heat ions during the thermalization process.

We do not have any specific astrophysical object in mind. From a theoretical perspective of non-relativistic shocks, we performed PIC simulations with relatively fast plasma flows, adopting upstream four-velocities of $u_0=0.2$ and $0.4$ to reduce computational cost.
The flow speeds of the Earth's bow shock is typically several hundred km/s, and the Alfv\'en Mach number seems to be in the range discussed in the manuscript.  For the case of supernova shocks, however, the typical shock speed is several thousand km/s, and the Alfv\'en Mach number is much higher than that discussed in this manuscript, and the additional heating by the excitation of Alfv\'en wave by the reflected ion beams seems to be less important.
Another application may be ultrafast outflows (UFOs) associated with supermassive black holes at galactic centers, which are known to reach velocities of 0.1c–0.3c \cite{Parker17,Gu25}. In this study, we focused on the energy partition between ions and electrons in non-relativistic shocks and in mildly relativistic shocks.

In this study, we compare ion and electron heating in two-dimensional shocks with that in one-dimensional shocks using particle-in-cell (PIC) simulations, and we \crd suggest \cbk the important role of Alfvén-wave dissipation and reconnection-driven ion heating in two-dimensional systems. However, in weakly magnetized shocks with high Alfvén Mach numbers, the free energy of the large-amplitude Alfvén waves—and thus the magnetic field forming the current sheets—is small, making reconnection-driven ion heating ineffective. Strong ion heating becomes significant only for relatively low–Alfvén-Mach-number shocks.

We have investigated plasma heating in two-dimensional parallel shocks. If fully three-dimensional shock simulations were performed, we expect that reconnection-driven energy dissipation would become even more effective than in two dimensions, because the reconnection structures would be more turbulent in fully three-dimensional magnetic fields \cite{Daughton11,Karimabadi13}.
For a three-dimensional guide-field magnetic reconnection with a pair plasma, it has bee shown that the nonthermal particle production is enhanced compared with that in two-dimensional system \cite{Hoshino24}.  Although we did not discuss the energy partitioning between thermal and nonthermal particles in this paper, this is another important problem, and will be investigated in future work.

In our simulation study, the ion–electron mass ratio is set to 100. As discussed in Appendix A, the ion-cyclotron resonance analyses remain essentially the same as those at the realistic mass ratio within the nonrelativistic approximation. However, it remains an open question whether the quantitative energy partitioning between ions and electrons during the nonlinear evolution of parametric instability, magnetic reconnection, and turbulent dissipation depends on the mass ratio.
Furthermore, our investigation of energy partitioning has so far focused on the case of parallel shocks, whereas quasi‑perpendicular shocks are likely to exhibit different partitioning characteristics. We have only begun to explore the potentially rich structures that govern energy partitioning in collisionless shocks. A systematic investigation over a broader range of shock parameters is therefore necessary.
\cbk
\begin{acknowledgments}
This work was supported by JSPS Grant-in-Aid for Scientific Research (KAKENHI) (Grant no. 24K06987).
The author (MH) would like to thank S. Totorica, G. T. Paz, and J. Niemiec for their valuable discussions.
This work used the computational resources of  the Fugaku supercomputer provided
by the RIKEN Center for Computational Science (Project ID: hp250048).
\end{acknowledgments}
\section*{DATA AVAILABILITY}
The data support the findings of this study are available from the corresponding author upon reasonable request.

\appendix
\section{Ion-cyclotron beam instability}
\begin{figure*}
\centering
\includegraphics[width=16cm]{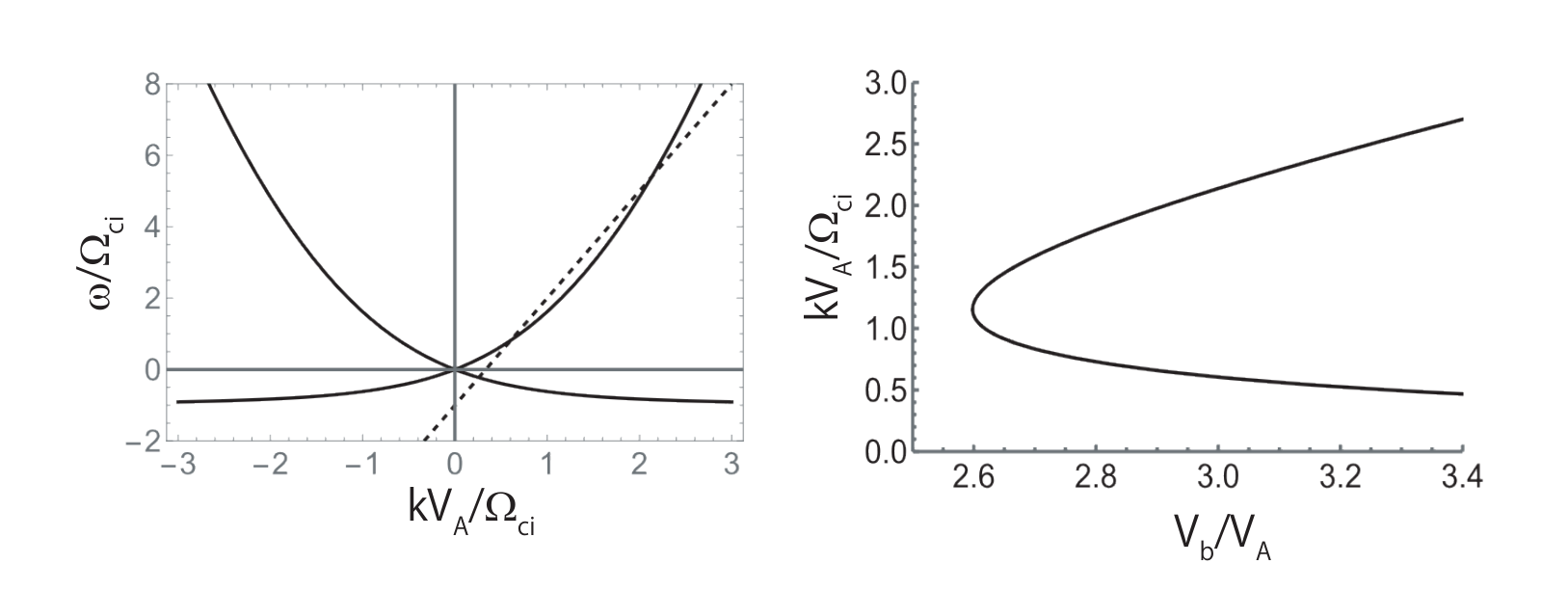}
\caption{(Left) Wave–particle interaction between ion beams and MHD waves. The dispersion relation in $\omega$–$k$ space is shown for the right-hand polarized whistler wave and the left-hand polarized ion–cyclotron/Alfv\'en wave. The dashed line indicates the ion-beam branch. (Right) Resonant wavenumbers $k c/\omega_{\rm pi}$ as a function of the beam speed $V_b/V_A$.}
\label{fig:ICB1}
\end{figure*}

A basic wave-particle interaction involved in the ion-beam cyclotron instability \cite{Gary81,Winske84,Hoshino85} is 
schematically illustrated in Figure \ref{fig:ICB1}.
The solid curve shows the dispersion relation for low-frequency MHD waves,
\begin{equation}
\omega^{2} = \left( 1 + \frac{\omega}{\Omega_{\rm ci}} \right) (k V_A)^2 ,
\end{equation}
where $\omega$, $k$, $V_A$, and $\Omega_{\rm ci}$ denote the wave frequency, wavenumber, Alfv\'en speed, and ion-cyclotron frequency, respectively \cite{Stix92}.
The right-hand polarized whistler mode corresponds to the region $\omega > 0$, whereas the 
left-hand polarized ion-cyclotron mode corresponds to $\omega < 0$. In the low-frequency limit $\omega \ll \Omega_{\rm ci}$, 
both right-hand and left-hand polarized modes are commonly referred to simply as Alfv\'en/MHD waves.
The dashed line represents the ion-beam mode,
\[
\omega = k V_b - \Omega_{\rm ci},
\]
with beam speed $V_b$. The intersection between the two branches indicates a possible point of momentum exchange between 
the wave and the beam plasma. When the beam speed exceeds approximately $2.6\,V_A$, three resonance points appear between the Alfv\'en-wave branch and the ion-beam branch. Under momentum conservation for both the wave and the beam plasma, 
however, only the right-hand polarized whistler waves with $\omega > 0$ and $k > 0$ can grow in time, whereas the 
left-hand polarized ion-cyclotron waves with $\omega < 0$ and $k > 0$ remain stable and cannot be excited \cite{Gary81,Winske84}.

Based on this simple picture of the ion-beam cyclotron instability, we now estimate the wavenumbers generated in the 
upstream region of the shock in our simulation. Using the simulation parameters of
$V_A/c \sim \Omega_{\rm ci}/\omega_{\rm pi} \sim 0.13$ and $V_b/c \sim 0.37$ discussed in Section II, we obtain 
$k(c/\omega_{\rm pi}) \sim 0.54$ in the upstream rest frame. After Lorentz transforming into the simulation frame, the 
corresponding value becomes $k'(c/\omega_{\rm pi}) \sim 0.60$. This estimate is consistent with the longer-wavelength mode observed in Figures~\ref{fig:stack1d} and \ref{fig:stack2d}.
The low-frequency whistler/MHD waves predominantly propagate parallel to the reflected ion beam, but their group velocity is slower than the upstream flow speed. Consequently, the waves are convected toward the downstream region and become 
compressed together with the plasma density. 

We briefly comment on additional important properties of the ion-cyclotron beam instability when finite ion and electron temperatures are included. The plasma dispersion relation is given by
\begin{equation}
  1 - \frac{k^2 c^2}{\omega^2}
  + \sum_j \frac{\omega_{pj}^2}{\omega^2} \zeta_j Z_0(\zeta_j^{\pm}) = 0,
\label{eq:ICBI}
\end{equation}
where $Z$ is the plasma dispersion function, 
$\zeta_j = (\omega - k v_{b,j})/(k v_{th,j})$, 
$\zeta_j^{\pm} = (\omega - k v_{d,j} \pm \Omega_{cj})/(k v_{th,j})$, 
and $v_{d,j}$, $v_{th,j}$, and 
$\omega_{pj} = \sqrt{4\pi n_j e^2/m_j}$ denote the drift velocity, thermal velocity, and plasma frequency of each species $j$, respectively.

\begin{figure*}
\centering
\includegraphics[width=16cm]{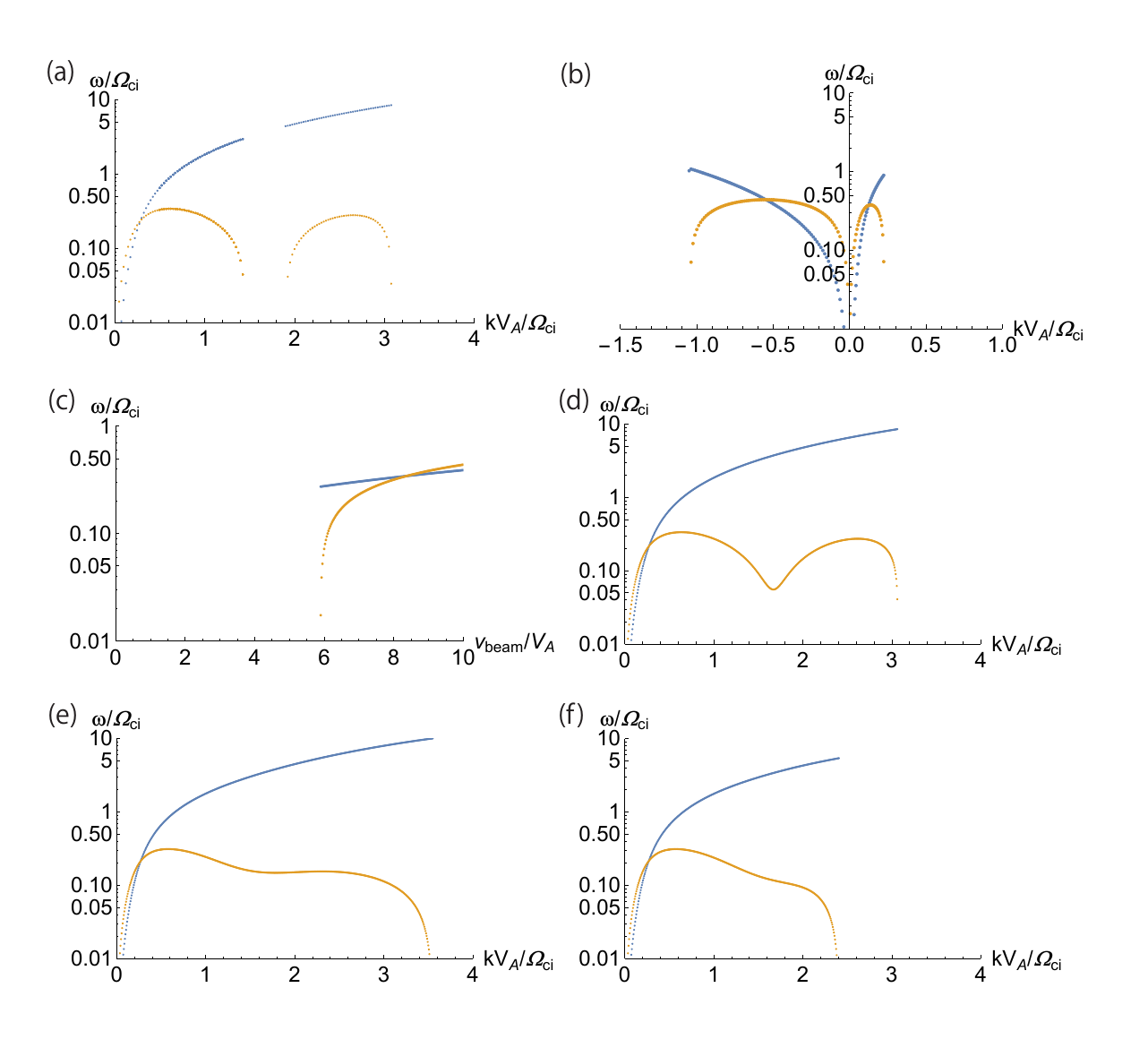}
\caption{Growth rate ($\mathrm{Im}(\omega)/\Omega_{\mathrm{ci}}$, orange) and real frequency 
($\mathrm{Re}(\omega)/\Omega_{\mathrm{ci}}$, blue) of the ion–cyclotron beam instability. 
Panels (a)–(c) show results obtained under the cold-plasma approximation. 
Panel (a) presents the resonant mode for a beam velocity $V_b/V_A = 3$, while panel (b) shows both the resonant and 
non-resonant modes for $V_b/V_A = 10$. Panel (c) illustrates the dependence of the non-resonant mode on the beam 
velocity $V_b$. Panels (d)–(f) include finite-temperature effects for $V_b/V_A = 3$. The reflected-ion temperature is fixed at $T_{ri} = m_i V_A^2/2$. In panel (d), the incoming ions and electrons are cold. In panel (e), the incoming-ion 
temperature is $T_{ii} = T_{ri}/10$ and the electron temperature is $T_e = T_{ri}$. In panel (f), the electron 
temperature is increased to $T_e = 3 T_{ri}$.}
\label{fig:ICB2}
\end{figure*}
In our analysis of whistler-wave generation in the shock region, we assume two ion components and a single electron component, and the subscript $j$ denotes $ri$, $ii$, and $e$ for the reflected ions, incoming ions, and hot electrons, respectively.
The instability is analyzed in the electron rest frame, and the two ion components satisfy the current-free condition $n_{ii} v_{d,ii} + n_{ri} v_{d,ri} = 0$.

Figure~\ref{fig:ICB2} shows the growth rate (orange) and real frequency (blue), both normalized by the ion-cyclotron 
frequency $\Omega_{\rm ci}$. Panels (a)–(c) correspond to the cold-plasma approximation for both ions and electrons, whereas panels (d)–(f) include finite-temperature effects. The beam (incoming) ion density is fixed at 10\%.

Figures~\ref{fig:ICB2}(a) and (b) show the results for beam/incoming drift velocities $v_{d,ii}/V_A = 3$ and $10$, 
respectively, as functions of the wavenumber $k V_A/\Omega_{\mathrm{ci}}$ normalized by the ion inertial length. For the lower beam velocity, only the resonant mode propagating parallel to the beam direction ($k > 0$) is excited, whereas for the higher beam velocity both resonant and non-resonant modes appear, generating waves in both directions. It is well known that the non-resonant mode can be excited when the bulk-flow energy density $n_{ii} v_{d,ii}^2$ is sufficiently large \cite{Winske84}. 
As shown in Figure~\ref{fig:ICB2}(c) as a function of the beam velocity for $k V_A/\Omega_{\rm ci}=0.5$, the non-resonant mode becomes unstable only when the beam/drift velocity satisfies $v_{d,ii}/V_A > 6$.

For higher beam velocity, only the long wavelength mode is excited. As shown in Figure~\ref{fig:ICB1}, the frequency of the higher frequency mode increases with increasing $v_{d,ii}$, eventually becoming much larger than the ion cyclotron frequency. In addition, when the full dispersion relation of the whistler mode is taken into account, no resonant point exists in this higher frequency range for sufficiently large $v_{d,ii}$.

Figures~\ref{fig:ICB2}(d)–(f) show cases with beam/incoming drift velocity $v_{d,ii}/V_A = 3$ and hot reflected ions with thermal velocity $v_{th,ri}/V_A = 1$, corresponding to a temperature $T_{ri} = m_i V_A^2/2$. In Figure~\ref{fig:ICB2}(d), the incoming ions and electrons remain cold, and the results are nearly identical to the cold reflected-ion case, except for a finite growth rate around $k V_A/\Omega_{\rm ci} \sim 1.5$. When the incoming ions are assumed to be warm with temperature $T_{ii} = T_{ri}/10$ and the electrons have temperature equal to that of the reflected ions ($T_e = T_{ri}$), the growth rate is shown in Figure~\ref{fig:ICB2}(e). Due to the velocity spread of the ion beams, slightly higher-frequency waves with $k V_A/\Omega_{\rm ci} < 3.5$ can be excited. In Figure~\ref{fig:ICB2}(f), a hotter electron temperature $T_e = 3 T_{ri}$ is assumed, and we find that whistler waves with shorter wavelengths are damped as the electron temperature increases, due to cyclotron damping by energetic electrons.

To understand the simulation result, we assumed the mass ratio of $m_i/m_e=100$ in the above analysis, but the behavior is basically same as that of the realistic mass ratio as long as the non-relativistic approximation. Note that by taking into account of the relativistic effect the cyclotron resonance condition for electrons can be given by 
\[
\omega = -k v_{th,e} + \Omega_{\mathrm{ce}}/\gamma_e,
\]
where $v_{th,e}$, $\Omega_{\mathrm{ce}}$, and $\gamma_e$ denote the characteristic velocity of energetic electrons, the electron cyclotron frequency, and the Lorentz factor, respectively. In this case, energetic electrons can more efficiently absorb high-frequency whistler waves due to the reduction of the effective cutoff frequency $\Omega_{ce}/\gamma_e$.

\section{Ion acoustic instability}
In addition to the whistler waves generated by the reflected ion beams, ion–acoustic waves may also be excited around the shock transition region \cite{Akimoto85}, because the plasma consists of counter-streaming ions and hot electrons. Ion–acoustic waves are generally strongly damped due to ion Landau damping, except in environments where the electron temperature is sufficiently high. In front of the shock, the ions consist of relatively cold incoming ions and hot back-streaming ions, and the corresponding electrons also consist of incoming and back-streaming populations with elevated temperatures. Because hot electrons escape from the shock-front region, the phase speed of the ion–acoustic wave becomes larger than the ion thermal velocity, allowing the counter-streaming ions to serve as a free-energy source for exciting ion–acoustic waves.

To examine the possibility of ion–acoustic wave generation, we briefly analyze the plasma dispersion relation,
\begin{equation}
1 - \sum_j \frac{\omega_{pj}^2}{k^2 v_{th,j}^2} Z_j^{\prime}(\zeta_j) = 0,
\label{eq:IAI}
\end{equation}
where $\zeta_j = (\omega - k v_{d,j})/(k v_{th,j})$ and 
$Z^{\prime}(\zeta_j) = -2 \left[ 1 + \zeta_j Z(\zeta_j) \right]$ is the derivative of the plasma dispersion function. For simplicity, we assume a single hot-electron component, while the ions consist of two components: a fast, colder population with drift velocity $v_{d,ii}$ and a slower, hotter population with drift velocity $v_{d,ri}$. The subscript $j$ denotes $ii$, $ri$, and $e$ for the incoming ions, reflected ions, and hot electrons, respectively. The analysis is performed in the electron rest frame under the ion current–free condition $n_{ii} v_{d,ii} + n_{ri} v_{d,ri} = 0$. 

As seen in the electron phase-space plots in Figures~\ref{fig:snap1d} and \ref{fig:snap2d}, the electrons are already heated in front of the shock. Therefore, we approximate them as a single hot-electron component. The qualitative behavior of the ion–acoustic instability does not change even if a two-component electron population is assumed.

\begin{figure*}
\centering
\includegraphics[width=16cm]{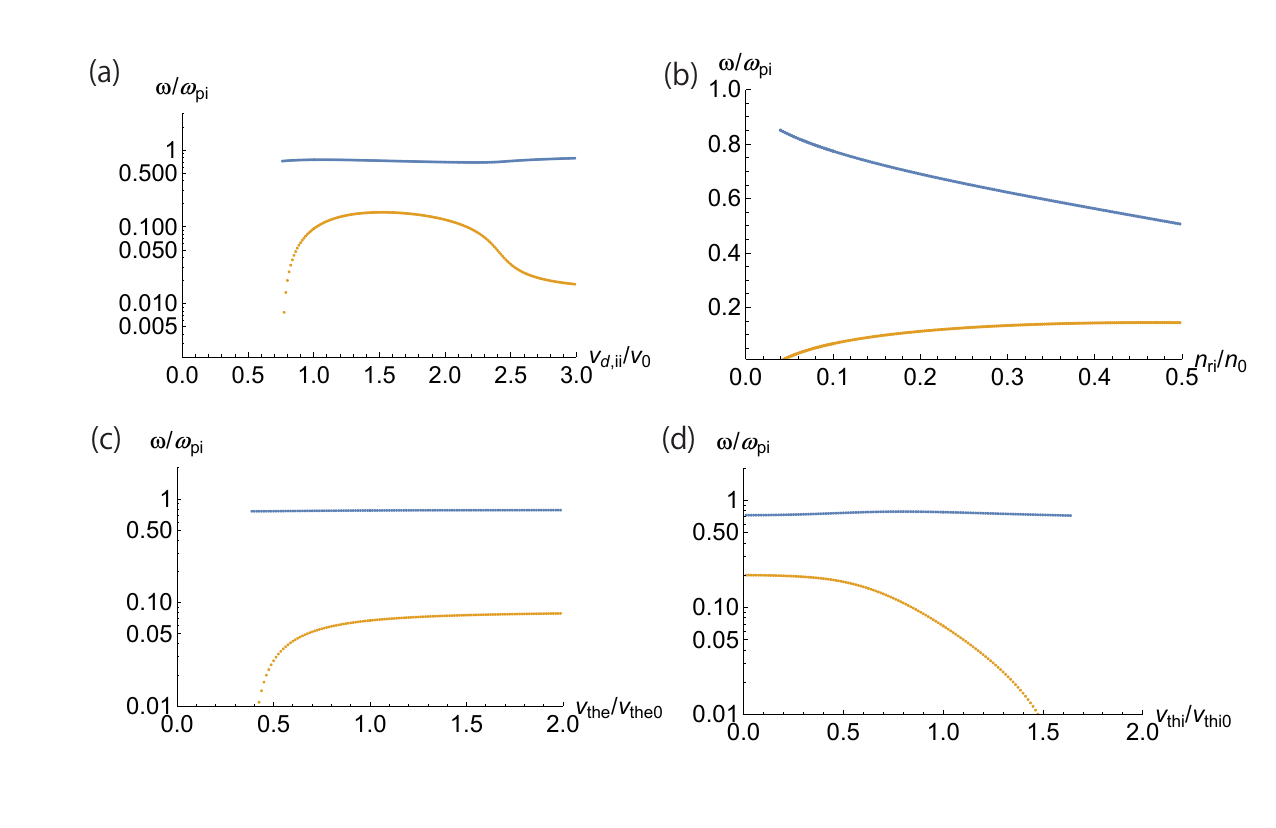}
\caption{Growth rate ($\mathrm{Im}(\omega)/\omega_{\rm pi}$, orange) and real frequency 
($\mathrm{Re}(\omega)/\omega_{\rm pi}$, blue) of the ion–acoustic instability. 
Panel (a) shows the dependence on the drift velocity of the incoming ion, where the normalization velocity $v_0$ corresponds to a typical incoming ion velocity in the shock upstream region. 
Panel (b) presents the dependence on the reflected ion density $n_{\rm ri}$ normalized by the total ion density $n_0=n_{\rm ri}+n_{\rm ii}$.
Panels (c) and (d) illustrate the effects of the electron temperature (electron thermal velocity $v_{the}$) and the 
reflected-ion temperature (reflected-ion thermal velocity $v_{thi}$), respectively.}
\label{fig:IAI}
\end{figure*}
Figure~\ref{fig:IAI} shows the growth rate (orange) and real frequency (blue) of the ion–acoustic instability. The growth rate and real frequency are normalized by the ion plasma frequency $\omega_{\rm pi}$.
We define the reference temperature as $T_0 = m_i v_0^2/3$, where $v_0$ is the upstream incoming ion velocity. To model the upstream shock conditions, we assume reflected and incoming ion temperatures of $T_{ri} = T_0/2$ and $T_{ii} = T_{ri}/10=T_0/20$, respectively, while the hot-electron temperature is set to $T_e = 10 T_0$. The density of the reflected ions is taken to be 10\% except for Figure~\ref{fig:IAI}(b). The wavenumber is set to be $k v_{d,ii}/\omega_{\rm pi}=1$, which is close to the maximum unstable wavenumber in the parameter range discussed here.

Figure~\ref{fig:IAI}(a) shows the growth rate (orange) and real frequency (blue) as functions of the incoming ion velocity and we find that the maximum growth rate occurs around $v_{d,ii}/v_0 \sim 1$. 
Figure~\ref{fig:IAI}(b) shows the dependence of the reflected ion density, with the incoming-ion drift velocity fixed at $v_{d,ii} = v_0$. The growth rate decreases as the reflected-ion density decreases, reflecting the reduction of free energy available to excite the ion–acoustic mode.

Figures~\ref{fig:IAI}(c) and (d) show the dependence on the electron thermal velocity and the reflected-ion thermal 
velocity, normalized by $v_{the0} = \sqrt{2 T_e/m_e} = \sqrt{20 T_0/m_e}$ and $v_{thi0} = \sqrt{2 T_{ri}/m_i} = \sqrt{T_0/m_i}$,
respectively.
The growth rate decreases as the electron thermal velocity decreases, due to enhanced ion Landau damping when the ion–acoustic phase velocity approaches the ion thermal velocity. For the same reason, the growth rate decreases as the ion thermal velocity increases.

These results demonstrate the fundamental behavior of ion–acoustic instability in a plasma composed of two ion components and a single hot electron component. The region immediately upstream of the shock corresponds to conditions where $v_{\mathrm{beam}}/v_0 \sim 1$ and both electron and ion temperatures remain relatively moderate. In this region, the reflected ions may possess sufficient free energy to excite the instability, even though the electron temperature is not necessarily high.  In contrast, in the region behind the shock front, both electrons and ions are substantially heated, and the counter-streaming ion velocity  becomes small. Although the ion drift as a free‑energy source might no longer be sufficient to drive the instability, the elevated electron temperature is favorable for its excitation. Overall, the results suggest that the excitation of the ion–acoustic wave is marginal.

\end{document}